\RequirePackage{rotating}
\documentclass[acmtosem]{acmart}

\usepackage{multirow} 
\AtBeginDocument{%
  }

\setcopyright{acmlicensed}
\copyrightyear{2026}
\acmJournal{TOSEM}
\acmYear{2026}
\acmDOI{XXXXXXX.XXXXXXX}
\acmISBN{978-1-4503-XXXX-X/2018/06}

\usepackage{framed}
\usepackage{xcolor}
\usepackage{subcaption}
\usepackage{rotating}
\usepackage{hyperref}
\usepackage[pagewise]{lineno}

\usepackage{tcolorbox}

\usepackage[scaled=0.92]{helvet}  
\usepackage{colortbl} 
\definecolor{warningborder}{HTML}{CCCCCC}
\definecolor{logbg}{HTML}{F8FAFC}
\definecolor{logborder}{HTML}{E2E8F0}
\definecolor{inlinebg}{HTML}{FFF0F2}
\definecolor{inlinetext}{HTML}{D63384}

\definecolor{diffbg}{HTML}{161B22}        
\definecolor{difflinegray}{HTML}{484F58}  
\definecolor{difftext}{HTML}{C9D1D9}      

\definecolor{diffkeyword}{HTML}{FF7B72}   
\definecolor{difffunc}{HTML}{D4A5F9}      
\definecolor{diffvar}{HTML}{79C0FF}       

\definecolor{diffaddbg}{HTML}{143525}     
\definecolor{diffaddnum}{HTML}{3FB950}    

\newcommand{\figurecode}[1]{\colorbox{inlinebg}{\textcolor{inlinetext}{\texttt{#1}}}}

\usepackage{draftwatermark}
\SetWatermarkText{ACCEPTED}
\SetWatermarkScale{0.7}
\SetWatermarkColor[gray]{0.9} 

\definecolor{lightgray}{gray}{0.9} 
\colorlet{shadecolor}{lightgray}
\newenvironment{highlightbox}
  {\begin{shaded}\setlength{\fboxrule}{1pt}\setlength{\fboxsep}{5pt}}
  {\end{shaded}}

\newcounter{finding}
\renewcommand{\thefinding}{Key Finding~\arabic{finding}:}
\newcommand{\finding}[1]{%
  \refstepcounter{finding}%
  \noindent\textbf{\thefinding} #1\par
}
\newcommand{\edits}[1]{\textcolor{black}{#1}}
\begin{document}
\title{Human Attention During Localization of Memory Bugs in C Programs}

\author{Emory Michaels}
\email{emory.michaels@nd.edu}
\orcid{0009-0001-5227-4930}
\affiliation{%
  \institution{Department of Computer Science and Engineering, University of Notre Dame}
  \city{Notre Dame}
  \state{Indiana}
  \country{USA}
}

\author{Robert Wallace}
\email{rwallac1@nd.edu}
\affiliation{%
  \institution{Department of Computer Science and Engineering, University of Notre Dame}
  \city{Notre Dame}
  \state{Indiana}
  \country{USA}}

\author{Matthew Robison}
\email{mrobison@nd.edu}
\affiliation{%
  \institution{Department of Psychology, University of Notre Dame}
  \city{Notre Dame}
  \state{Indiana}
  \country{USA}
}

\author{Yu Huang}
\email{yu.huang@vanderbilt.edu}
\affiliation{%
 \institution{Department of Computer Science, Vanderbilt University}
 \city{Nashville}
 \state{Tennessee}
 \country{USA}
 }

\author{Collin McMillan}
\email{cmc@nd.edu}
\affiliation{%
  \institution{Department of Computer Science and Engineering, University of Notre Dame}
  \city{Notre Dame}
  \state{Indiana}
  \country{USA}
  }


\begin{abstract}
This paper presents a study of human visual attention during localization of memory bugs in C.  Human visual attention refers to the mechanical processes by which we selectively process and prioritize information.  Visual attention is important to study because it is central to what information people (who are sighted) use to solve a particular problem.  Meanwhile, memory bugs are among the most common types of bugs in C programs that manifest as a variety of program faults.  In this paper, we study human visual attention while people attempt to locate memory bugs in code.  We recruited 21 programmers to locate between one and eight memory bugs in three C programs for 1.5--2 hours each.  In total we collected observations of 31 hours of programmer effort.  The bugs in our study cover memory leaks, overflows, and double frees, which are among the most common memory bugs. We analyze the task outcomes in terms of success rate and related factors, patterns of visual attention overall such as what lines and functions are read, and finally we explore differences of visual attention patterns during success versus failure cases.  
\end{abstract}

\ccsdesc[500]{Software and its engineering~Software development techniques}
\ccsdesc[300]{Social and professional topics~Computing education}
\ccsdesc[500]{Human-centered computing~Empirical studies in HCI}

\keywords{bug localization, memory bugs, human attention, visual attention, eye-tracking}

\received{10 December 2025}

\maketitle


\section{Introduction}
\label{sec:intro}

This paper presents a study of human attention during localization of C memory bugs.  Human visual attention refers to the mechanical processes by which we selectively process and prioritize information~\cite{carrasco2011visual}.  Human attention can be measured through eye-tracking, interface logs, and activity surveys.  It is important to study because beliefs about human visual attention influence decisions such as personnel allocation, user interface design, and even the architecture of machine neural networks~\cite{novak2024eye,baharum2024enhancing,cheng2012eye,zhang2024eyetrans,zhang2025enhancing, pourhosein2025unveiling,kiseleva2020study}.  Human attention has also long been academically interesting for ``greater'' philosophical reasons as a window into the working of our own minds~\cite{mole2025encylopedia}.

Meanwhile, memory bugs are one of the most common programming errors in C programs~\cite{van2012memory,chromium_memory_safety}.  Memory bugs are behind aberrant software behavior such as excessive memory use, slowdowns, security vulnerabilities, and crashes.  Some memory bug behaviors such as buffer overflows and segmentation faults are so common that they have become tropes in C programming lore~\cite{durumeric2014matter,orman2003morris}.  
Yet while often the cause of a memory bug is relatively small---a missing {\small \texttt{free()}}, an array length miscalculation, etc.---memory bugs can be difficult for programmers to find.  Complex programs can have many pointers active at a time and memory leaks can be hard to observe as they accumulate slowly in long-running programs.  This difficulty has made memory bugs a risk in C programs and a major research focus~\cite{van2012memory,oorschot2023memory}. 

Cotroneo~\emph{et al.}~\cite{cotroneo2016bugs} point out that ``use of uninitialized data, buffer overflows, and memory leaks are preponderant'' as causes of bug behavior related to memory.  Part of what makes these difficult to find is that they can lead to intermittent problems or challenges with vague non-functional requirements like responsiveness.  For example it may be possible to use uninitialized data for a long time before it causes a problem, even if it is not strictly legal.  Memory leaks may manifest as slow degradation in user experience rather than a diagnosable error message.  Better understanding how people find these types of bugs could help address the lion's share of challenges in localizing C memory bugs.

In this paper, we conduct an experiment in which programmers read C code and find memory bugs.  We extracted eight bug reports from three large, popular C programs.  Each bug is related to the three dominant memory bug types described in the previous paragraph or was a common bug type in the open-source C repositories.  We recruited 21 programmers; in the study each programmer worked for 1.5 to 2 hours to find as many of the eight presented bugs as possible.  During the study we used eye-tracking hardware to observe each programmer's visual attention, while also tracking other interactions.  We measured programmer performance such as success rate in finding bugs and time required.  We also used a minimally-invasive attentional state survey for a self-assessment of attentiveness. \textcolor{black}{Together, these data characterize visual attention during localization of bugs in C programs.}

We found that: 1) people spend 75\% of their fixations on just 25\% of the functions they fixate on; 2) even unsuccessful participants fixate on the correct buggy line, but fail to recognize it; 3) successful bug locators have higher regression rates and average shorter distances between fixations; 4) successful bug locators revisit and/or spend more time on the buggy line, while unsuccessful participants do not return and/or spend less time on it.

\section{Background and Related Work}

This section discusses studies of human attention, studies of the process of bug localization, and the use of eye-tracking in program comprehension research.

\subsection{Studies of Human Attention}

Eye-tracking has been used extensively to study human attention. For example, computational models of eye movement during reading have been developed to predict how a person’s eyes will move over a body of text~\cite{reichle2003ez, engbert2005swift}. These models highlight that eye movement across text can be predicted by textual features, which cause people to move from word to word, but also pause and reread text to ensure understanding. Such models have also been applied to non-reading tasks, such as visual search~\cite{reichle2012using}. Collectively, the modeling efforts have revealed that there are systematic patterns of eye movements that people follow when reading or performing other visual tasks. Consequently, eye-tracking has also been employed to examine when and why people deviate from those stereotypical patterns, thereby identifying the oculomotor correlates of mindlessness (or mind-wandering) during reading and other tasks. Specifically, multiple studies have found systematic differences in eye movements when people are reading attentively versus inattentively~\cite{reichle2010eye, foulsham2013mind, frank2015validating, bixler2016automatic}. Specifically, when people are mind-wandering while reading, their eye movements are less sensitive to the content and complexity of the text, moving over the text unpredictably and unsystematically. 

These phenomena have also been studied in other contexts, such as watching video lectures to simulate online coursework~\cite{zhang2020wandering}, listening to audiobooks~\cite{faber2020eye}, and viewing images of scenes~\cite{faber2020eye, krasich2018gaze}. Faber ~\emph{et al}.~\cite{faber2020eye} have highlighted that inattentiveness is not always accompanied by the same patterns of eye movements across various tasks, as they require differences in the focality/dispersion of visual attention. Therefore, the task itself must be considered when examining such differences. The present study will thus seek to identify systematic oculomotor patterns when people are localizing bugs in code in an effort to understand how people do and do not detect bugs, and what their eyes are doing when that happens.

\vspace{-7mm}
\subsection{Eye-tracking in Program Comprehension}

Eye-tracking technology has been a bedrock of program comprehension research for at least two decades and recent systematic surveys have led to recommended best practices in using this technology~\cite{sharafi2015systematic, Sharafi2020PracticalGuide}.  Over time simultaneous decreases in hardware costs and quality increases have expanded access to eye-tracking equipment and the number of papers using it~\cite{braw2023integrating}.

Program comprehension itself has a long history as research discovering how people read and understand source code~\cite{cornelissen2009systematic, dominic2020program, maalej2014comprehension, schroter2017comprehending, stapleton2020human}.  Sometimes formalized as the ``concept assignment problem''~\cite{biggerstaff1993concept}, the idea is that reading code is the process of interpreting the low-level implementation details in code as high-level concepts about program behavior.  To study how developers engage with these low-level details, researchers often use eye-tracking to examine the mechanics of visual attention. One fundamental component of this analysis is fixations, when the eyes briefly pause on a specific point to allow the brain to process visual information~\cite{grabinger2025cookbook}. Analyzing fixations provides insight into how developers allocate attention during code reading.  For example, Aschwanden and Crosby~\cite{aschwanden2006code} observed how programmers' eyes moved while examining a snippet of code featuring a loop and equations.  Bednarik and Tukiainen~\cite{bednarik2006eye} focused on scanpaths (i.e., sequences of eye movements) as participants tried to make sense of two small programs.  Their findings revealed that programmers’ mental strategies sharpen over time, as indicated by fewer shifts in visual focus. Peitek ~\emph{et al.} showed that knowledge guides the eyes of proficient programmers~\cite{peitek2022correlates}. Later, Rodeghero~\emph{et al.}~\cite{rodeghero2014improving} turned their attention to how developers write documentation, using this information to inform advances in automated summarization tools. Followup examination of these results extracted and categorized different eye-tracking movement strategies~\cite{rodeghero2015empirical, rodeghero2015eye}.  Abid~\emph{et al.}~\cite{abid2019developer} expanded on Rodeghero’s study~\cite{rodeghero2014improving}, incorporating real-world project contexts and a realistic IDE interface (better interfaces are possible with newer eye-tracking devices and infrastructure).  Similar findings have been observed in related areas such as reading UML diagrams, with some diagram formats leading to more thorough program comprehension than others~\cite{gueheneuc2006taupe, Sharif:2010:ELD:1796177.1796642}.

A few papers have sought to combine programmer visual attention with neural network-based models to predict human behavior or change machine behavior to be more like human behavior.  For example, Karas~\emph{et al.}~\cite{karas2024tale} compared eye movement pattern strategies when reading code for different tasks, then used these results to design novel neural network architectures to do those tasks~\cite{zhang2024eyetrans}.  Likewise Bansal~\emph{et al.}~\cite{bansal2023scanpath} modeled eye movements as scanpaths and augmented the attention mechanism in a neural network to improve code documentation generation~\cite{bansal2023human}.  The idea of combining human with AI attention is starting to take root outside of software engineering, with preliminary evidence showing performance improvements in various tasks such as image recognition and text generation~\cite{fong2018using, lopez2024seeing}.

\subsection{Studies of Bug Localization}
\label{subsec:background_bug}

Bug localization is the task of finding the area of code that causes a bug behavior~\cite{saha2013improving, thung2014buglocalizer, Obaidellah2018}.  Bug localization is a popular research target because understanding how people find bugs could lead to better automated tools to assist people.  Studies of bug localization have revealed strategies roughly falling into either static or dynamic approaches.  Static strategies are those reading code itself while dynamic approaches execute that code.  Dynamic approaches are well-studied and include tactics like program logging, assertions, breakpoints, and profiling~\cite{wong2016survey}.  Static approaches are less well-studied, though they are a common starting point for programmers, especially when the code exists in a large codebase with which the programmer is not yet familiar.  This situation is common when onboarding new staff, for example~\cite{dominic2020onboarding, ju2021case}.  This paper is primarily concerned with static approaches and specifically the mechanical process of attention during bug localization.  Related papers include eye-tracking studies such as by Bednarik and Tukiainen~\cite{bednarik2008temporal} who analyze the effect of experience level, finding that people with more experience read more areas of code but in less depth than novices. \textcolor{black}{Tablatin found that low-performing students focused more on debugging easier errors than high-performing students~\cite{tablatin2023visual}.} Similarly to results from program comprehension studies, Hejmady
and Narayanan analyzed scanpaths during debugging tasks and found that better programmers have fewer
attentional shifts~\cite{hejmady2012visual}. Research indicates that eye movements differ between debugging code and reading for other tasks~\cite{turner2014eye}.  One difference is that programmers search code more often in some tasks than others~\cite{sharafi2020eyes}.  Given that differences exist in different tasks, this paper focuses on the narrow task of localization of memory bugs in C programs to avoid an assumption that movement patterns for other tasks generalize to this specific task.

\vspace{-3mm}
\section{Experimental Design}

Our experiments center around observing what people pay attention to while trying to statically locate memory bugs in C programs.  We study micro- and macro-level attention.  By ``micro level'' we mean mechanical details of attention such as fixation and regression counts.  By ``macro level'' we mean what types of information people are reading and in what order.  By ``statically'' we mean by just reading the code only, without executing it.  By ``locate bugs'' we mean identify the line of code where the bug behavior originates.  We define these terms in more detail throughout this section.

\vspace{-4mm}
\subsection{Research Questions}
Our research objective is to understand the visual attention of programmers localizing bugs in C programs and whether these patterns differ between when participants successfully locate the bug versus cases when the bug is not identified. 
We ask the following Research Questions (RQs):
\vspace{-1mm}
\begin{description}
    \item[~~RQ1] What are the outcomes of the task completion in terms of success rates, confidence, and other measures of human performance?
    \item[~~RQ2] What patterns of visual attention do people follow during the bug localization tasks?
    \item[~~RQ3] What patterns of visual attention are associated with successful versus failed task completion outcomes?
\end{description}

The rationale for RQ1 is to understand the degree to which people are able to complete the tasks as we present them in the experimental setting.  An experiment with a diverse set of participant skill and task difficulty would generally lead to a diverse set of task outcomes.  Therefore we start our analysis of attention by reporting the landscape of task outcomes.  Next, our rationale behind RQ2 is to understand the visual attention that people follow during bug localization of C memory bugs.  We seek to understand both micro- and macro-level attention.  While there have been studies of attention for other tasks, we focus specifically on memory bugs in C programs.  Finally, our rationale for RQ3 is to connect RQ1 and RQ2 results: we seek to understand if there are patterns of attention that are more or less likely to lead to success.  These results are important because they could lead to educational or tool support interventions.

\subsection{Procedure Overview}

Our experimental procedure is a single-group cohort~\cite{Kazdin2016}, ``Quadrant II'' study described by Stol and Fitzgerald~\cite{stol2020guidelines}.  What this means is that we focus on a single-group common characteristic (people proficient in C programming) in a controlled environment (as opposed to ``Quadrant I'' field studies, for example).  We chose a single-group design because we do not seek to distinguish human factors such as age or gender, and we chose a controlled environment to reduce experimental variables.  We 1) curate a list of bugs that is representative of typical memory bugs in C programs, 2) recruit programmers proficient in C to read bug reports and source code from that list, and 3) observe what the people do in a controlled environment. 

The participant's objective for each bug was to read the bug report, read the source code for the project containing the bug behavior, and find the line(s) that the participant believes are the cause of the bug.  We did not ask the participant to actually fix the bug or run the program as our research goal is to understand the process of static bug localization.

We curated a pool of eight bugs, and we recruited 21 people.  We present each person \textcolor{black}{with} a maximum of eight bugs to find over 1.5 to 2 hours.  Three bugs were the same for all participants and three to five rotated from our pool.  To reduce experimental variables, we conducted the study one participant at a time. Each participant was physically present in a quiet computer laboratory setting and used the same hardware and software.  Note that because we are primarily interested in patterns of attention, we avoided distracting participants from the computer.  We chose a minimally-invasive design where an author of this paper was available for experimental setup at all times, but this person did not intervene or ask the participant questions in an e.g. think-aloud form. The scope of this paper is a quantitative analysis of the eye-tracking data.

\begin{table}[h]
    \caption{Statistics for C libraries from which we selected bugs.}
    \centering
    \begin{tabular}{lrrr}
        \toprule
        Library    & $\sim$LoC & Domain & Contributors \\
        \midrule
        OpenSSL   & 1.4m  & security      & 980        \\
        redis     & 250k  & database      & 734        \\
        sway      & 50k   & graphical     & 475        \\
        \bottomrule
    \end{tabular}
    \label{tab:c_libraries_stats}
\end{table}

\subsection{Subject Bug Reports}
\label{sec:bugs}

We selected a total of eight bug reports from three C libraries/programs.  \textcolor{black}{Our process for selecting the bugs was 1) Search github for C repositories. 2) Search the repositories for resolved memory bugs. 3) Evaluate each bug. To evaluate each bug, we consider bug type and bug complexity.} Our goal is for these bug reports to be representative of typical memory bugs in C.  For a definition of ``typical'' we turned to Cotroneo~\emph{et al.}~\cite{cotroneo2016bugs} who studied characteristics of bugs and found that memory bugs are almost always one of three types: uninitialized data, buffer overflows, and memory leaks.  We use bugs of the buffer overflow and memory leak type, but used double frees in lieu of uninitialized data due to their prevalence in our datasets.  Bugs also have a range of difficulty and severity.  At the same time we must balance diversity of bug selection with the practical reality of human studies, as has been pointed out in retrospective studies e.g. by Siegmund~\emph{et al.}~\cite{siegmund2015views}.  Participants will learn about a program and seek to reuse that knowledge on other bugs, they will have different prior knowledge, have time limits, and will fatigue. 

Table~\ref{tab:c_libraries_stats} shows the three C libraries/programs we used.  These programs have a range of total size and domain.  Total size is important because finding bugs in larger programs may be more difficult.  Domain is important because coding style may differ depending on domain~\cite{cox2009programming}.  We then selected between one and three bugs from the bug report repository for each program.  
\textcolor{black}{We used comments from the bug report conversation, area of code that needed to be changed to fix the bug, and our own judgement to choose bugs of different difficulties.}
\textcolor{black}{We eliminated bugs that require too much project knowledge to solve in 30 minutes.}  We also ensured that all bugs were double-frees, buffer overflows, or memory leaks. Table ~\ref{tab:bug_complexity} shows additional information about each bug. See Section~\ref{sec:threats} for a discussion of the risks and limitations of our choice of bugs.  Also see our reproducibility package linked in Section~\ref{sec:conclusion}.
\begin{table}[h]
    \color{black}
    \caption{\color{black}Issue and code complexity information for each bug. Column ``Program'' is the github repository. Column ``Issue'' is the issue number in the project's bug tracking system, included here for reproducibility. Column ``Type'' is the memory bug type. Column ``Codename'' is the name we used in our experiments and in this paper's text for more memorable record-keeping. Column ``File / Function'' contains either the name of the function that contains the chosen bug or the name of the file that contains the chosen bug. Some bugs have multiple functions listed because there were alternate answers that we accepted. Column ``NLOC'' is the number of non-comment lines of code contained in the file/function. ``CCN'' is the cyclomatic complexity of the file/function. Column ``Token'' is the number of tokens contained in the file/function. ``Param'' is the number of parameters in the file/function. ``Length'' is the total number of lines in the file/function.}
    \centering
    \setlength{\tabcolsep}{3pt} 
    \begin{tabular}{lllllrrrrr}
        \toprule
        \textcolor{black}{\textbf{Codename}} & \textbf{Program} & \textbf{Issue} & \textbf{Type}  & \textbf{File / Function} & \textbf{NLOC} & \textbf{CCN} & \textbf{Token} & \textbf{Param} & \textbf{Length} \\
        \midrule
        \textcolor{black}{\textbf{ladybug}} & sway & 6861 & leak &  seat\_device\_destroy & 12 & 2 & 69 & 1 & 13 \\
        & & & & \textit{seat.c file total} & 1323 & 343 & 8387 & 158 & 1497 \\
        \midrule
        \textcolor{black}{\textbf{stonefly}} & redis & 8299 & overflow & rdbWriteRaw & 5 & 3 & 37 & 3 & 5 \\
        & & & & rdbSaveLzfBlob & 17 & 5 & 146 & 4 & 24 \\
        & & & & \textit{rdb.c file total} & 2133 & 626 & 15008 & 144 & 2653 \\
        \midrule
        \textcolor{black}{\textbf{hornet}} & sway & 3577 & double free &  execute\_command & 94 & 24 & 677 & 3 & 107 \\
        & & & & config\_command & 65 & 23 & 519 & 2 & 83 \\
        & & & & config\_subcommand & 17 & 3 & 116 & 4 & 18 \\
        & & & & config\_commands\_command & 69 & 13 & 445 & 1 & 80 \\
        & & & & \textit{commands.c file total} & 397 & 102 & 2633 & 26 & 451 \\
        \midrule
        \textcolor{black}{\textbf{mantis}} & redis & 4527 & overflow &  stringmatchlen & 119 & 36 & 561 & 5 & 121 \\
        & & & & \textit{util.c file total} & 562 & 164 & 3946 & 37 & 696 \\
        \midrule
        \textcolor{black}{\textbf{firefly}} & openssl & 24892 & leak &  show\_digests & 22 & 7 & 168 & 2 & 28 \\
        & & & & \textit{dgst.c file total} & 487 & 137 & 2948 & 18 & 551 \\
        \midrule
        \textcolor{black}{\textbf{silverfish}} & redis & 9800 & leak &  PingReceiver & 6 & 1 & 76 & 5 & 6 \\
        & & & & \textit{hellocluster.c total} & 51 & 10 & 428 & 19 & 66 \\
        \midrule
        \textcolor{black}{\textbf{spider}} & sway & 3300 & double free &  swaynag\_destroy & 48 & 12 & 329 & 1 & 58 \\
        & & & & \textit{swaynag.c file total} & 355 & 96 & 2564 & 57 & 385 \\
        \midrule
        \textcolor{black}{\textbf{weevil}} & openssl & 20278 & double free &  OSSL\_HTTP\_get & 78 & 15 & 451 & 13 & 82 \\
        & & & & \textit{http\_client.c file total} & 1087 & 365 & 6707 & 130 & 1322 \\
        \bottomrule
    \end{tabular}
    \label{tab:bug_complexity}
\end{table}

For each bug, we \textcolor{black}{read} 1) the bug report, 2) follow-up communications among programmers and users, 3) the version of the code containing the bug, and 4) the patch fix showing lines changed to fix the bug. \textcolor{black}{Then, we identified which parts of the bug report make the bug easier or harder to solve or explicitly mention the bug or fix. Our goal was to change bug reports as little as possible, but we do remove parts of the bug report that explicitly answered the questions we asked participants. For some of the bug reports, we added a hint to help the participants locate the correct directory or file because it was especially difficult to find without being familiar with the repository's structure. Figure~\ref{fig:ladybug_report} shows an example of a bug report adapted from issue \#6861 in the \texttt{sway} repository. This is the bug report for \texttt{ladybug}.}

\begin{figure}[ht]
    \centering
    \begin{minipage}{0.95\textwidth} 
        \fontfamily{phv}\selectfont 
        
        \setlength{\parskip}{1em}
        \setlength{\parindent}{0pt}
        
        {\LARGE\bfseries ladybug - sway[sway d6f2799]\par}
        
        {\large\bfseries Project Synopsis:\par}
        sway is an i3-compatible Wayland compositor
        
        \begin{tcolorbox}[
            colback=white, colframe=warningborder, sharp corners, boxrule=0.8pt,
            top=8pt, bottom=8pt, left=10pt, right=10pt
        ]
            \small\textbf{Please make sure you have CALIBRATED and STARTED TRACKING before starting!}
        \end{tcolorbox}
        
        \begin{center}
            \texttt{-------------------------START OF BUG REPORT-------------------------}
        \end{center}
        
        {\large\bfseries Title: sway\_switch is never destroyed\par}
        
        \begin{tcolorbox}[
            colback=logbg, colframe=logborder, arc=4pt, boxrule=1pt,
            top=10pt, bottom=10pt, left=12pt, right=12pt
        ]
        \begin{flushleft}
            \ttfamily\small
            Direct leak of 280 byte(s) in 7 object(s) allocated from:\\
            \hspace*{1.5em}\#0 0x7f18aa588fb9 in \_\_interceptor\_calloc /usr/\allowbreak src/\allowbreak debug/\allowbreak gcc/\allowbreak libsanitizer/\allowbreak asan/\allowbreak asan\_malloc\_linux.cpp:154\\
            \hspace*{1.5em}\#1 0x5562660d724c in sway\_switch\_create ../sway/sway/input/switch.c:9\\
            \hspace*{1.5em}\#2 0x5562660d724c in seat\_configure\_switch ../sway/sway/input/seat.c:815\\
            \hspace*{1.5em}\#3 0x5562660d724c in seat\_configure\_device ../sway/sway/input/seat.c:881
        \end{flushleft}
        \end{tcolorbox}
        
        \figurecode{sway\_switch} devices are created but never destroyed on teardown.
        
        \begin{center}
            \texttt{-------------------------END OF BUG REPORT-------------------------}
        \end{center}
    \end{minipage}
    
    \vspace{1em} 
    \caption{\textcolor{black}{This is the bug report for \texttt{ladybug}. This report is adapted from issue \#6861 in the sway repository.}}
    \Description{Here is the text of the bug report: START OF BUG REPORT. Title: sway_switch is never destroyed. Direct leak of 280 byte(s) in 7 object(s) allocated from:
    #0 0x7f18aa588fb9 in __interceptor_calloc /usr/src/debug/gcc/libsanitizer/asan/asan_malloc_linux.cpp:154
    #1 0x5562660d724c in sway_switch_create ../sway/sway/input/switch.c:9
    #2 0x5562660d724c in seat_configure_switch ../sway/sway/input/seat.c:815
    #3 0x5562660d724c in seat_configure_device ../sway/sway/input/seat.c:881. sway_switch devices are created but never destroyed on teardown. END OF BUG REPORT}
    \label{fig:ladybug_report}
\end{figure}


\subsection{Study Participants}

We recruited 21 participants for this study.  The participants had between 3 and 13 years of programming experience.  Eleven were graduate students, nine were undergraduates, and one was a professional software engineer.  Eight of the graduate students worked in the software engineering industry before coming to graduate school. Seven of the undergraduates had professional experience at software engineering internships. The undergraduates were recommended as strong C programmers by faculty, though these faculty were not authors of this study and this study was not connected to any coursework (per IRB requirements).  We compensated participants at the fully-loaded local market rate of 100\,USD/hr.

\vspace{-2mm}
\subsection{Procedure Details}

The details of our study procedure boil down to: 1) each participant visited the laboratory in person at a scheduled time, one at a time, 2) the participant sat at the eye-tracking computer to become familiar with the settings and review informed consent documents, and 3) read and locate C memory bugs for a maximum of 120 minutes.  We gave each participant a maximum of eight bug reports to read \textcolor{black}{and informed participants that they have a maximum of 30 minutes per bug}.  Some of the bug reports were the same for everyone, while we rotated others (as a balance between diversity of samples and measuring consistency of participant behavior). \textcolor{black}{Every participant saw the bugs: \texttt{ladybug} \textcolor{black}{(a \texttt{sway} bug)}, \texttt{stonefly} \textcolor{black}{(a \texttt{redis} bug)}, and \texttt{spider} \textcolor{black}{(a \texttt{sway} bug)}. Two bugs (\texttt{ladybug} and \texttt{stonefly}) were always presented first as a baseline, and the third bug (\texttt{spider}) was included among the remaining 4–6 reports.\footnote{\edits{This strategy establishes a baseline and allows future research into bug presentation order. Note that because participants completed different total numbers of bugs, some did not reach the third bug (\texttt{spider}).}} The rest of the reports were rotated from the remaining pool (see Appendix 5.1.2).}

\begin{figure}[t]
    \centering
    \includegraphics[width=1.0\linewidth]{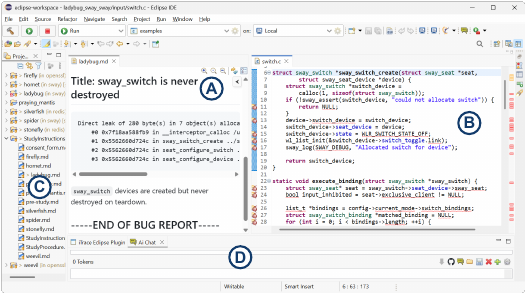}
    \caption{The interface presented to participants.  Area A is the bug report, area B is the code, area C is the file explorer, and area D is a text chat window with GPT-4o for AI help.}
    \Description{This image shows the Eclipse IDE setup presented to participants. On the far left of the screen is area C which is the file explorer. To the right of the file explorer, there are 3 panels. The upper portion of the screen is divided into 2 panels: areas A and B. The lower portion of the screen is area D: the AI chat window. Area A has the \texttt{ladybug} \textcolor{black}{(a \texttt{sway} bug)}bug report showing on it. Area B has the file switch.c showing. The AI chat window is blank.}
    \label{fig:interface}
\end{figure}

An example image of the interface participants saw is in Figure~\ref{fig:interface}.  To reduce experimental variables and ensure accurate eye-tracking, we did not allow the participants to navigate away from the interface; we instead made necessary information available within the interface.  The interface itself is based on the Eclipse 4.33.0 IDE with iTrace 0.2.0~\cite{itracetool}.  We set the IDE to have four areas, shown in Figure~\ref{fig:interface}.  Area A is information about the bug including the bug report.  We showed only the portion of the report explaining the bug symptoms. We did not allow the participants to see followup in which the exact bug location may be revealed.  Area A is also where participants wrote their answers.  

Area B is a typical code editing window.  Area C is a typical file navigation area.  For simplicity we had each bug localization task as a separate folder; clicking each folder would load the bug information into area A and reveal the relevant program's code.  We ensured each participant worked on only one bug at a time, though we interrupted the participant if  \textcolor{black}{they} had not located a bug within thirty minutes and moved the participant to the next bug.

We regulated access to outside resources via a chat window with GPT-4o using the RemainAI~\cite{remainAIChat} Eclipse plugin (Area D in Figure~\ref{fig:interface}).  This window was necessary because: 1) we did not allow focus to be changed from the IDE interface, 2) we did not want participants to find the bug's solution from a solved bug report online, 3) in practice AI help is quickly becoming an important part of the programming process so participants expect this support, and 4) the plugin allows us to capture queries that people type.  We set the plugin so that it would not automatically have access to the code and we asked participants to ask general coding help questions instead of details about the program.  This restriction was necessary because GPT-4o could have memorized the bug report information during training.

We created a minimally-invasive popup window to appear every five minutes.  This window asked ``In the few seconds before this screen appeared, what were you thinking about?'' with options to answer:

{
\begin{itemize}
    \item I was focused on the task.
    \item I was thinking about my performance on the task or how long it is taking.
    \item I was thinking about things unrelated to the task (e.g., daydreaming).
    \item I was distracted by sights or sounds in my environment.
    \item My mind was blank.
\end{itemize}
}

The purpose of this window is to measure self-reported attention level. This method is used in cognitive psychology to generate momentary subjective assessments of attentional states~\cite{weinstein2018mind, robison2019examining, smallwood2015science}. Responses to these screens can be used to examine which signals, both in terms of overt behavioral responses and physiological indices, co-occur with and precede moments of attentiveness versus inattentiveness. Similarly, in the present study, we sought to understand whether certain behaviors/eye-tracking metrics would be associated with self-reported inattentiveness by participants.

When a participant was ready with an answer to a bug report,  \textcolor{black}{they}  answered \textcolor{black}{``cause''} and \textcolor{black}{``location''} questions.  To answer \textcolor{black}{``location''},  \textcolor{black}{they}  entered the line number location into a specified area in the bug report window (Figure~\ref{fig:interface}, area A).  The participant could enter a range of lines if  \textcolor{black}{they}  could not decide or believed more than one line was involved.  To answer \textcolor{black}{``cause''}, the participant entered a brief description in English of what  \textcolor{black}{they} believed the problem to be. \textcolor{black}{For both ``cause'' and ``location'' responses, participants rated confidence in each answer on a 1–5 scale (1: least confident, 5: most confident). Participants also rated perceived difficulty on a 1–5 scale (1: least difficult, 5: most difficult).}


\subsection{Data Processing}
\label{sec:dataprocessing}

We collected seven types of information:

{
\begin{itemize}
    \item Eye-tracking data (Tobii Pro Fusion 120Hz, iTrace 0.2.0)
    \item Screen video capture (1080p, 30 FPS)
    \item Communication with AI help (GPT-4o)
    \item Self-reported attention level (five minute intervals)
    \item Self-reported bug difficulty and confidence
    \item Bug cause answered by participant (\textcolor{black}{``cause''})
    \item Bug location answered by participant (\textcolor{black}{``location''})
\end{itemize}
}

Two authors (with 5 years and 14 years of teaching and grading experience respectively) graded the participants' answers on a scale of 1-5 where 1 was very inaccurate and 5 was highly accurate. Attempts without answers received 0. A score of highly-accurate meant that the reported bug cause or location was identical to the one in the official issue or fixed by the official pull request or of equal plausibility as determined by the authors.  For example, for \texttt{ladybug} \textcolor{black}{(a \texttt{sway} bug)}, the location of the cause of the bug is line \texttt{48} (before fix): \texttt{free(seat\_device);} because when \texttt{seat\_device} is freed, access to \texttt{seat\_device->switch\_device} is lost, and the allocated memory for \texttt{seat\_device->switch\_device} is leaked. Notice that the bug cause location is \textit{not} the fix location.
\begin{table}[h]
    \centering
    \caption{Rubric for grading \textcolor{black}{``cause''} questions.}
    \label{tab:what_rubric}
    \renewcommand{\arraystretch}{1.2}

    \begin{tabular}{c p{7.2cm}}
        \toprule
        \textbf{Score} & \textbf{Bug Identification and Explanation} \\
        \midrule
        5 & Correctly identifies and explains bug with details. \\
        4 & Correctly identifies bug and explains with some details. Given more time, would have provided more details. On the right path. \\
        3 & Partly correct in identifying bug and explaining, or somewhat vague. Possibly some conceptual or detail errors. \\
        2 & Mostly incorrect in identifying bug and explaining, or extremely vague with no details. \\
        1 & Missed basics of bug or completely incorrect with little to no details. \\
        \bottomrule
    \end{tabular}
\end{table}

\begin{table}[H]
    \centering
    \caption{Rubric for grading \textcolor{black}{``location''} questions.}
    \label{tab:where_rubric}
    \renewcommand{\arraystretch}{1.2}

    \begin{tabular}{c p{7.2cm}}
        \toprule
        \textbf{Score} & \textbf{Bug Location and Thought Process} \\
        \midrule
        5 & Correct line, backed up by correct thought process. \\
        4 & Given more time, could have narrowed down to the correct line, but was close (correct function or call stack). \\
        3 & Mostly correct understanding of problem, some misunderstandings, limited idea where to look. \\
        2 & Incorrect understanding of problem or not much beyond bug report, but provided a location that follows logic of incorrect understanding. \\
        1 & Incorrect understanding of problem and vague location, random guess. \\
        \bottomrule
    \end{tabular}
\end{table}

To minimize bias, we applied the rubrics shown in Tables~\ref{tab:what_rubric}~and~\ref{tab:where_rubric} to decide scores. These rubrics reflect knowledge from related literature, the teaching experience of the authors, and the range of answers from participants. These rubrics are holistic, meaning that graders form an overall impression of the answer rather than assigning points for each criterion independently~\cite{fitzgerald2013we}. The response characteristics that we looked for are similar to the requirements of the ``isolation process'' for debugging described in Hylton ~\emph{et al.}~\cite{hylton2023board}, namely that the most complete answers have where the cause of the error is, the reasoning about which part of the code causes the error, and reasoning about how the error could have happened. Or, in other words, the most complete bug localizers know where, what, \textit{and} why.  \textcolor{black}{Coherence is important and should be a fundamental category in rubrics [25]. In order to receive a ``location’’ accuracy score of ``2’’ instead of ``1,’’ a participant’s ``location’’ answer must follow the logic of their ``cause’’ answer.} 

To resolve disagreements, we followed a discussion-based approach because it is recommended by related work for situations with a low tolerance for imperfect grading~\cite{rieser2011reinforcement,wood2018detecting,oortwijn2021interrater}. We followed the tested procedure from Oortwijn \emph{et al.}~\cite{oortwijn2021interrater}, which involves five steps. Step 1 is setting up the annotation task and guidelines. In our study, we use the rubrics in Tables \ref{tab:what_rubric} and \ref{tab:where_rubric}, and each annotator annotates all participant responses. Step 2 is inter-rater conceptual alignment. During this step, we decide on the meaning of terms in our study like ``the cause of the bug'' and ``the location of the bug.'' These terms also had to be defined and explained to our participants, so we used the definitions and examples from the participant instructions. Step 3 is individual annotation when annotators score each response independently. Step 4 is annotation comparison. \textcolor{black}{The pre-discussion quadratically weighted Cohen's kappa coefficient for \textcolor{black}{``Cause''} accuracy was 0.853, and for \textcolor{black}{``Location''} accuracy it was 0.846.} Step 5 is disagreement resolution.  As noted in Oortwijn \emph{et al.}~\cite{oortwijn2021interrater}, annotators often find that disagreements are due to ``task or guideline unclarity.'' When this happens, annotators should clarify the guidelines. During this step, we discussed every disagreement and came to a consensus on what the final score should be, resulting in one final score for each response. \textcolor{black}{See Figure~\ref{fig:grading_consensus_examples} for two examples of this resolution process.}

\vspace{-3mm}
\begin{figure}[htbp]
    \centering
    \small
    
    \begin{tcolorbox}[colback=gray!3, colframe=gray!60, boxsep=2pt, left=6pt, right=6pt, top=6pt, bottom=6pt, title=\textbf{Example \#1: P18 Silverfish}, fonttitle=\bfseries]
        \textbf{Participant’s Cause Answer:} I am not sure about the cause for this bug. I have looked into the code in \texttt{hellocluster.c}, and the function \texttt{PingallCommand\_RedisCommand} and \texttt{PingReceiver} since the memory leak happens at the callback. My guess of the bug is that the program may create some memory in the heap for the string or anything to callback, but does not release it.
        
        \smallskip
        \textbf{Participant’s Location Answer:} My guess is from line 76 - 81 related to the \texttt{PingReceive} and other functions related to this, but I am not sure.
        
        \smallskip
        \hrule
        \smallskip
        \smallskip
        
        \textbf{Grader \#1 Score:} 4 \quad | \quad \textbf{Grader \#2 Score:} 3 \quad | \quad \textbf{Consensus Score:} \textbf{4}
        
        \smallskip
        \textbf{Consensus Discussion:} Before discussing, grader 2 wrote: \emph{“not a bad guess but doesn't quite get there, could be convinced this is a 4 since possibly would have made it with more time.”} We ultimately agreed to give this participant a 4 instead of a 3 because there are no errors in their answer. The description for a “Cause” score of 3 says, “Partly correct in identifying bug and explaining, or somewhat vague. Possibly some conceptual or detail errors.” This participant seems more unsure than vague, and they are definitely on the right track, even providing a range of line numbers in the “Location” answer that includes the correct line. We believe this participant was on the right track and could have provided more details given more time which is the description for a “Cause” score of 4.
    \end{tcolorbox}
    
    \vspace{0.3em}
    
    \begin{tcolorbox}[colback=gray!3, colframe=gray!60, boxsep=2pt, left=6pt, right=6pt, top=6pt, bottom=6pt, title=\textbf{Example \#2: P2 Hornet}, fonttitle=\bfseries]
        \textbf{Participant’s Cause Answer:} I guess it's because \texttt{free\_workspace\_config} be called to free the memory for the second time when there's invalid input, i.e., the fourth parameter is not a number. The first free might be Line 79 in \texttt{workspace.c}.
        
        \smallskip
        \textbf{Participant’s Location Answer:} In Line 77 in \texttt{workspace.c}, before calling the \texttt{strtol} function, we need to check if it's a valid string that can be converted as a int value amount.
        
        \smallskip
        \hrule
        \smallskip
        \smallskip
        
        \textbf{Grader \#1 Score:} 4 \quad | \quad \textbf{Grader \#2 Score:} 5 \quad | \quad \textbf{Consensus Score:} \textbf{4}
        
        \smallskip
        \textbf{Consensus Discussion:} Originally, Grader \#2 gave this participant a 5 because they saw that the participant had many details of the bug correct including the location of where to insert a fix. However, the participant was incorrect about \texttt{free\_workspace\_config}, and so we decided that a score of 4 was more appropriate.
    \end{tcolorbox}
    \vspace{-4mm}
    \caption{\textcolor{black}{Examples of Step 5: disagreement resolution for grading of participant answers.}}
    \label{fig:grading_consensus_examples}
\end{figure}

\vspace{-3mm}
To process and analyze the eye-tracking data, we used the ``Map Tokens'' and ``Generate Fixations'' tools in iTrace Toolkit V0.2.2 to extract fixations, lines, and tokens~\cite{behler2023itracetoolkit}. The inputs to the iTrace Toolkit are: iTrace-Core data, iTrace Eclipse plugin data, and srcML archive files~\cite{collard2013srcML}. The fixation settings for generating fixations were: Fixation Filter: IVT~\cite{andersson2017one}, and the default settings for IVT: Velocity Threshold: 50, Duration (ms): 80. The iTrace Toolkit outputs a database containing fixation and gaze information. Then, we reviewed the screen recordings for each task and noted the x and y coordinates of each area of the screen for each portion of the session (Figure~\ref{fig:interface}, areas A-D). These coordinates varied for each task because we allowed each participant to configure their screen. We build on scripts from Wallace ~\emph{et al.}~\cite{Wallace2025Programmer}, and the inputs are: databases from iTrace Toolkit, the \textcolor{black}{``location''} and \textcolor{black}{``cause''} scores, screen region coordinates, and the self-reported confidence and difficulty scores. 

\edits{In our analysis, we use empirical metrics collected during the study: participant accuracy and participant-reported difficulty. We use these participant-driven outcomes because they are based on direct experimental results and previous literature uses participant performance as a proxy for difficulty~\cite{lin2015tracking,gould1975some}. Conversely, we do not utilize static code complexity metrics (such as those in Table~\ref{tab:bug_complexity}) or our own initial task assessments to quantify difficulty in the analysis. Our initial assessments were used solely during task curation to help select a diverse range of bugs for the study. Furthermore, we do not use static code complexity metrics because a task's true difficulty depends on many interacting factors, including the clarity of the bug report, function pointer relationships, and project file structure, for example. Because our study was not designed to isolate these specific independent variables, we do not evaluate how these individual factors impact difficulty, though we encourage future studies to do so.}

\subsection{Threats to Validity}
\label{sec:threats}

Our study has several threats to validity. One threat is that the selected bugs may not fully represent all C memory bugs. We mitigate this risk by choosing bugs as described in Section \ref{sec:bugs}. \textcolor{black}{We chose to focus on a subset of types of bugs to balance diversity and consistency. We acknowledge that other types of memory bugs like buffer underflows and use-after-frees were not present in our study and future work should study more types of bugs.}
A second threat is the varying detail in GitHub bug reports. To address this, we reviewed each report and made minor edits to ensure answering the \textcolor{black}{``cause''} and ``\textcolor{black}{location}'' questions was non-trivial.
Another threat is that while most of the participants have professional experience, most participants are currently graduate or undergraduate students, who may not fully represent industry professionals. To mitigate this, we selected only those with C experience and accepted only highly recommended undergraduates. \textcolor{black}{We acknowledge the potential mixed effect of participant experience and project domain.}
\textcolor{black}{The time limit and popups may have affected participants in different ways. Table~\ref{tab:popup} shows that participants reported ``thinking about my performance on the task or how long it is taking'' in 6.06\% of the popups. In the post-study survey, only one participant mentioned the popups saying, ``answering what I was thinking about influenced me to focus much more, since I felt I was being held accountable.'' To reduce the impact of popup interruptions, participants read the popup choices before starting the task. This means that participants did not have to read the answer choices every time the popup appeared. Instead, they could answer quickly because the answer choices were the same every time. While the time limit may impact some participants' performance on some tasks, 92\% of the tasks were completed in less than 28 minutes. This means that most participants chose not to use all of their time. In the cases where participants used more than 28 minutes (n=7), only 2 tasks were rated above a 3 on Location Accuracy. In the other 5 cases, participants did not seem to be on the right track after 30 minutes, so it is unclear if more time would have impacted their performance.}
Another threat is that some participants wore glasses and participants were not reminded to keep their heads still. However, the iTrace toolkit is designed to handle these concerns by using a conservative approach to identifying gazes and fixations. 

To address eye-tracker drift~\cite{Sharafi2020PracticalGuide} and occasional disconnections, we recalibrated the eye-tracker between tasks and after any interruptions.
An additional threat is that screen recordings often froze mid-session, affecting our ability to retrieve screen region coordinates and AI usage data. However, participants rarely changed window layouts after initial setup, and we used eye-tracking data to confirm AI usage based on the amount of time the participant looked at the AI window.
\textcolor{black}{Though participants could use the search tool, they could not use IDE features like ``jump to definition'' to help them with the task. This may have made the tasks more challenging by forcing participants to do more manual navigation.}
Also, the eye-tracker was not tracking during Participant 11's or Participant 18's \texttt{ladybug}  \textcolor{black}{(\texttt{sway})} task, so these tasks are not included in eye-tracking metrics.
Finally, Participant 11 did not complete the \texttt{firefly} \textcolor{black}{(\texttt{openssl})} task and Participant 13 did not complete the \texttt{mantis} \textcolor{black}{(\texttt{redis})} task. They did not answer the \textcolor{black}{``cause''} and ``\textcolor{black}{location}'' questions within 30 minutes. Since we cannot infer their understanding, they received a score of 0, and we do not include this data when comparing successful versus unsuccessful tasks.

\begin{figure*}[h]
\centering
\renewcommand{\arraystretch}{1}
\begin{tabular}{ccc}
\includegraphics[width=0.3\linewidth]{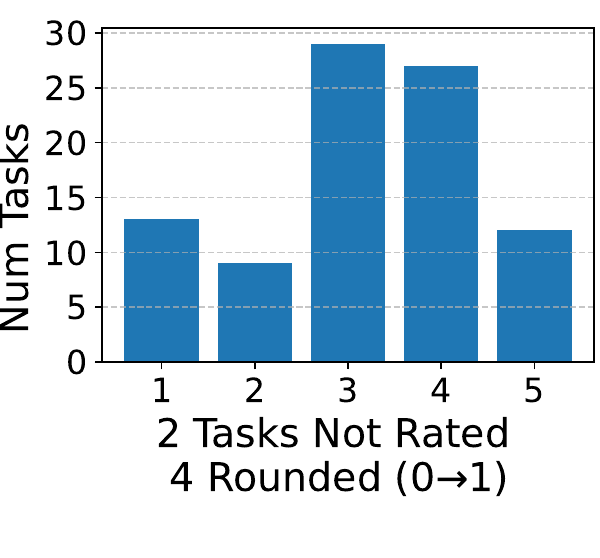} & \includegraphics[width=0.3\linewidth]{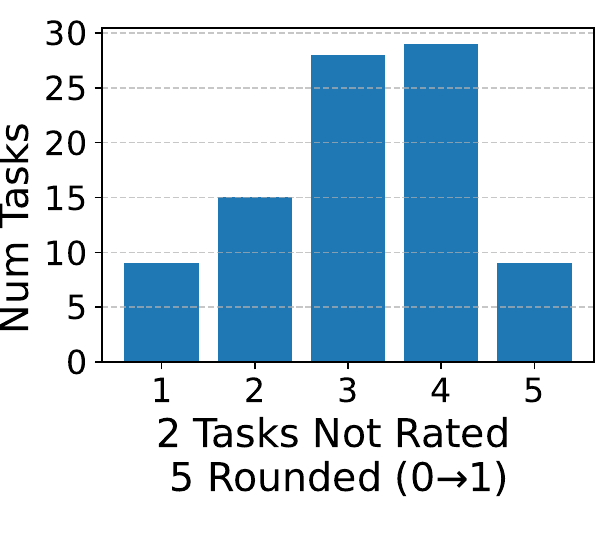} & \includegraphics[width=0.3\linewidth]{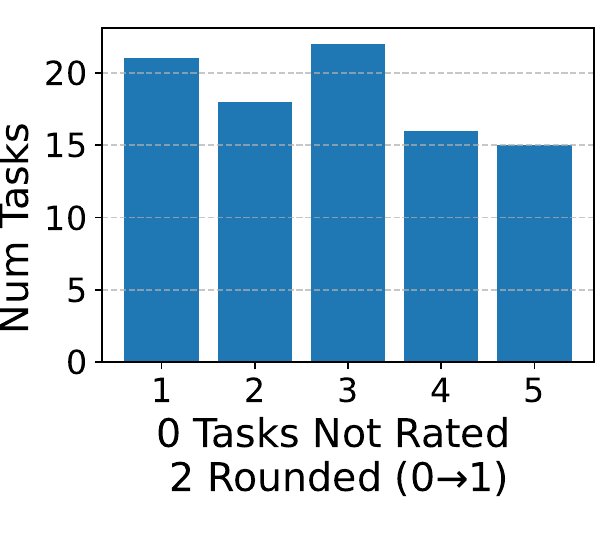} \\
(a) \textcolor{black}{``Cause''} Confidence & (b) \textcolor{black}{``Cause''} Difficulty & (c) \textcolor{black}{``Cause''} Accuracy \\
& & \\
\includegraphics[width=0.3\linewidth]{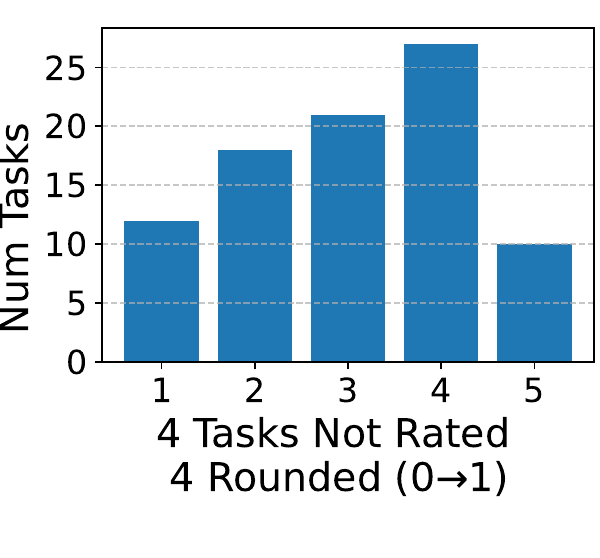} & \includegraphics[width=0.3\linewidth]{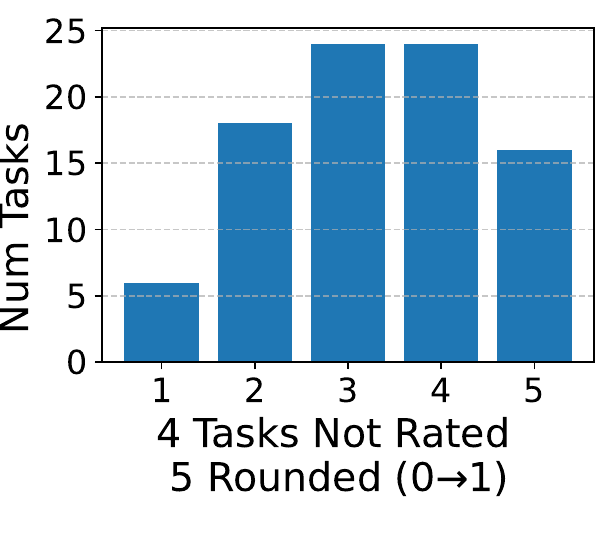} & \includegraphics[width=0.3\linewidth]{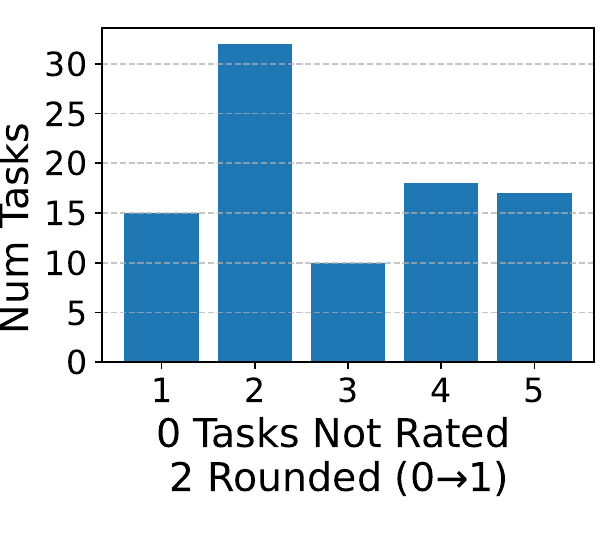} \\
(d) \textcolor{black}{``Location''} Confidence & (e) \textcolor{black}{``Location''} Difficulty & (f) \textcolor{black}{``Location''} Accuracy \\
  &  & 
\end{tabular}
\caption{Histograms of confidence, difficulty, and accuracy scores from all participants and tasks.  Confidence and difficulty are self-reported by participants.  Accuracy is graded on a rubric by the authors of this paper.  \textcolor{black}{``Cause''} refers to the bug cause.  \textcolor{black}{``Location''} refers to the bug location in code.}
\Description{This figure has 6 subfigures. They are in 2 rows of 3 and labeled a through f. a) shows a histogram of the number of tasks where participants rated their confidence in their answers to the \textcolor{black}{``cause''} question. b) shows a histogram of the number of tasks where participants rated the difficulty of answering the \textcolor{black}{``cause''} question. c) shows a histogram of the number of tasks that earned a \textcolor{black}{``cause''} accuracy score between 1 and 5. d) shows a histogram of the number of tasks where participants rated their confidence in their answers to the \textcolor{black}{``location''} question. e) shows a histogram of the number of tasks where participants rated the difficulty of answering the \textcolor{black}{``location''} question. f) shows a histogram of the number of tasks that earned a \textcolor{black}{``location''} accuracy score between 1 and 5.}
\label{fig:scores}
\end{figure*}

\section{Experimental Results}
\subsection{RQ1: Task Outcomes}

In this section we present our experimental results as answers to RQ1-3 and supporting evidence.

Overall, we found around a fifth of the \textcolor{black}{responses had completely correct reasoning and location reports.} 
About 18\% of the graded accuracy for \textcolor{black}{``location''} questions were 5, shown in Figure~\ref{fig:scores}(f), which corresponds to a finding of ``correct'' (Table~\ref{tab:where_rubric}).  Also, only around 16\% of the graded accuracy for \textcolor{black}{``cause''} questions were 5, shown in Figure~\ref{fig:scores}(c), which corresponds to a finding of ``correctly'' identifying the bug (Table~\ref{tab:what_rubric}).  However, these numbers rise to around 38\% if we include answers clearly on the right track, only limited by time (i.e. accuracy scores of 4 or higher).  These results show the difficulty of bug localization given low overall success, yet still point to the tractability of these bugs by people, considering that about half were at least on the right track despite only reading the bugs for a few minutes (49\% of \textcolor{black}{``location''} accuracy scores were 3 or higher).  Still, a minority did not show evidence that the problem would have been solved even given much more time --- for only around 16\% of tasks was there complete failure to find the correct location (score of 0 or 1 for \textcolor{black}{``location''} accuracy) and 23\% for describing the problem (score of 0 or 1 for \textcolor{black}{``cause''} accuracy), including total guesses or bug descriptions demonstrating key misunderstandings.


\begin{figure}[htbp]
    \centering
    \begin{minipage}{0.95\textwidth}
        \ttfamily\small 
        
        \begin{tcolorbox}[
            colback=diffbg, 
            colframe=diffbg, 
            arc=4pt, 
            boxrule=0pt,
            top=10pt, bottom=10pt, left=10pt, right=10pt
        ]
        \setlength{\tabcolsep}{4pt} 
        \renewcommand{\arraystretch}{1.1} 
        
        \begin{tabular}{l l l l}
            \textcolor{difflinegray}{37} & \textcolor{difflinegray}{37} & & \textcolor{diffkeyword}{static void} \textcolor{difffunc}{seat\_device\_destroy}\textcolor{difftext}{(}\textcolor{diffkeyword}{struct} \textcolor{difftext}{sway\_seat\_device} \textcolor{diffvar}{*}\textcolor{difftext}{seat\_device}\textcolor{difftext}{) \{} \\
            \textcolor{difflinegray}{38} & \textcolor{difflinegray}{38} & & \quad \textcolor{diffkeyword}{if} \textcolor{difftext}{(!}\textcolor{difftext}{seat\_device}\textcolor{difftext}{) \{} \\
            \textcolor{difflinegray}{39} & \textcolor{difflinegray}{39} & & \qquad \textcolor{diffkeyword}{return}\textcolor{difftext}{;} \\
            \textcolor{difflinegray}{40} & \textcolor{difflinegray}{40} & & \quad \textcolor{difftext}{\}} \\
            \textcolor{difflinegray}{41} & \textcolor{difflinegray}{41} & & \\
            \textcolor{difflinegray}{42} & \textcolor{difflinegray}{42} & & \quad \textcolor{difffunc}{sway\_keyboard\_destroy}\textcolor{difftext}{(}\textcolor{difftext}{seat\_device}\textcolor{diffvar}{->}\textcolor{diffvar}{keyboard}\textcolor{difftext}{);} \\
            \textcolor{difflinegray}{43} & \textcolor{difflinegray}{43} & & \quad \textcolor{difffunc}{sway\_tablet\_destroy}\textcolor{difftext}{(}\textcolor{difftext}{seat\_device}\textcolor{diffvar}{->}\textcolor{diffvar}{tablet}\textcolor{difftext}{);} \\
            \textcolor{difflinegray}{44} & \textcolor{difflinegray}{44} & & \quad \textcolor{difffunc}{sway\_tablet\_pad\_destroy}\textcolor{difftext}{(}\textcolor{difftext}{seat\_device}\textcolor{diffvar}{->}\textcolor{diffvar}{tablet\_pad}\textcolor{difftext}{);} \\
            
            \rowcolor{diffaddbg}
            \textcolor{diffaddnum}{\ } & \textcolor{diffaddnum}{45} & \textcolor{diffaddnum}{+} & \quad \textcolor{difffunc}{sway\_switch\_destroy}\textcolor{difftext}{(}\textcolor{difftext}{seat\_device}\textcolor{diffvar}{->}\textcolor{diffvar}{switch\_device}\textcolor{difftext}{);} \\
            
            \textcolor{difflinegray}{45} & \textcolor{difflinegray}{46} & & \quad \textcolor{difffunc}{wlr\_cursor\_detach\_input\_device}\textcolor{difftext}{(}\textcolor{difftext}{seat\_device}\textcolor{diffvar}{->}\textcolor{diffvar}{sway\_seat}\textcolor{diffvar}{->}\textcolor{diffvar}{cursor}\textcolor{diffvar}{->}\textcolor{diffvar}{cursor}\textcolor{difftext}{,} \\
            \textcolor{difflinegray}{46} & \textcolor{difflinegray}{47} & & \qquad \textcolor{difftext}{seat\_device}\textcolor{diffvar}{->}\textcolor{diffvar}{input\_device}\textcolor{diffvar}{->}\textcolor{diffvar}{wlr\_device}\textcolor{difftext}{);} \\
            \textcolor{difflinegray}{47} & \textcolor{difflinegray}{48} & & \quad \textcolor{difffunc}{wl\_list\_remove}\textcolor{difftext}{(}\textcolor{diffvar}{\&}\textcolor{difftext}{seat\_device}\textcolor{diffvar}{->}\textcolor{diffvar}{link}\textcolor{difftext}{);} \\
            \textcolor{difflinegray}{48} & \textcolor{difflinegray}{49} & & \quad \textcolor{difffunc}{free}\textcolor{difftext}{(}\textcolor{difftext}{seat\_device}\textcolor{difftext}{);} \\
            \textcolor{difflinegray}{49} & \textcolor{difflinegray}{50} & & \textcolor{difftext}{\}}
        \end{tabular}
        \end{tcolorbox}
    \end{minipage}
    
    \vspace{1em}
   \caption{Official fix for \texttt{ladybug} \textcolor{black}{(a \texttt{sway} bug)}. See Figure \ref{fig:ladybug_report} for the  \texttt{ladybug} bug report. Expected bug cause location is line 48 (before fix): \texttt{free(seat\_device);}.}
    \Description{This shows a code snippet from a github pull request. The function shown is seat_device_destroy. There is an inserted line calling sway_switch_destroy at line 45. At line 48 which becomes line 49 after the insertion is the call to free(seat_device).}
    \label{fig:ladybug_fix}
\end{figure}

To illustrate the range of participant answers, consider the bug \texttt{ladybug}, a memory leak of a variable called \texttt{seat\_device->switch\_device} \textcolor{black}{in the \texttt{sway} project}.
See Figures \ref{fig:ladybug_report} and \ref{fig:ladybug_fix} for more information about \texttt{ladybug}.
An example of an answer that received a \textcolor{black}{``cause''} score of 1: 
\begin{quote}
``The bug is that the calloc function is making a recursive call to the function \texttt{sway\_switch\_create} and is sending the wrong parameters. We are allocating possibly more memory than needed. The parameters are wrong, need to redefine before the call and do not send it the same function.''
\end{quote} 
This answer received a \textcolor{black}{``cause''} score of 1 because it is not related to the memory leak bug. An example of an answer that received a \textcolor{black}{``cause''} score of 3: 
\begin{quote}
``The cause of the bug is a calloc call that is not properly freed. The memory allocation that should be freed occurs on line 9.'' 
\end{quote}
This answer received a score of 3 because it is incomplete. While the participant does identify which memory is leaked, the participant does not identify where or why. An example of an answer that received a \textcolor{black}{``cause''} score of 5: 
\begin{quote}
``In file \texttt{seat.c}, line no 881, \texttt{seat\_configure\_switch} function to configure switch. In line 815, it creates a switch, if it does not exists. To do so, it calls \texttt{sway\_switch\_create method} from \texttt{switch.c} file. In this file, at line no 9, it allocates memory for the switch. However, when \texttt{seat\_device\_destroy method} is called to destroy the \texttt{seat\_device in seat.c} file, in this method from line 37 to 49, the allocated switch is not destroyed by calling \texttt{sway\_switch\_destroy} method from \texttt{switch.c} file. It can cause memory leak issue.''
\end{quote}
This answer received a score of 5 because it identifies the cause of the bug by explaining why the memory leak occurs.

\begin{table}[H]
    \caption{Pearson correlation coefficient matrix. Significant values (except diagonal) at $\alpha = 0.050$ are in bold.}
    \centering
    \begin{tabular}{llllllllll}
        \toprule
        & & \multicolumn{3}{c}{\textcolor{black}{Cause}} & \multicolumn{3}{c}{\textcolor{black}{Location}} \\
        \cmidrule(lr){3-5} \cmidrule(lr){6-8}
        &  & Confidence & Difficulty & Accuracy & Confidence & Difficulty & Accuracy \\ \cline{3-8}
        \multirow{3}{*}{\rotatebox[origin=c]{90}{\textcolor{black}{Cause}}} 
        & \multicolumn{1}{l}{Confidence}  & 1.000  & \textbf{-0.740} &  \multicolumn{1}{l}{\textbf{0.291}}  &  \textbf{0.542}  & \textbf{-0.452}  &  \textbf{0.272}  \\
        & \multicolumn{1}{l}{Difficulty}  & \textbf{-0.740} &  1.000  &  \multicolumn{1}{l}{-0.183}  & \textbf{-0.452}  &  \textbf{0.556}  & -0.174  \\
        & \multicolumn{1}{l}{Accuracy}    &  \textbf{0.291}  &  -0.183  &  \multicolumn{1}{l}{1.000}  & \textbf{0.376}  & \textbf{-0.339}  & \textbf{0.930}  \\ \cline{3-8}
        \multirow{3}{*}{\rotatebox[origin=c]{90}{\textcolor{black}{Location}}} 
        & \multicolumn{1}{l}{Confidence}  &  \textbf{0.542}  & \textbf{-0.452}  & \multicolumn{1}{l}{\textbf{0.376}}  &  1.000  & \textbf{-0.643}  & \textbf{0.324}  \\
        & \multicolumn{1}{l}{Difficulty}  & \textbf{-0.452}  &  \textbf{0.556}  & \multicolumn{1}{l}{\textbf{-0.339}}  & \textbf{-0.643}  &  1.000  & \textbf{-0.280}  \\
        & \multicolumn{1}{l}{Accuracy}    &  \textbf{0.272}  & -0.174  & \multicolumn{1}{l}{\textbf{0.930}}  & \textbf{0.324}  & \textbf{-0.280}  &  1.000  \\
        \bottomrule
    \end{tabular}
   
    \label{tab:correlation}
\end{table}

Table~\ref{tab:correlation} shows Pearson correlations among all scores for all tasks and participants, with significant t-test results in bold.  
Notably the self-reported difficulty for \textcolor{black}{``cause''} questions is not correlated to the rubric-graded accuracy for either question. 
 Still, confidence and difficulty scores are correlated, implying that participants are answering earnestly (random answers would tend to lead to uncorrelated responses).  
Other correlations include: 1) \textcolor{black}{``location''} accuracy to \textcolor{black}{``cause''} accuracy, which is not surprising considering that a correct rationale likely leads to a correct bug location, 2) \textcolor{black}{``location''} accuracy correlates to \textcolor{black}{``location''} confidence and \textcolor{black}{``location''} difficulty, implying that people are better judges of their performance of locating a bug, a finding supported by 3) \textcolor{black}{``location''} confidence is correlated with \textcolor{black}{``cause''} accuracy, in that cases where people are confident in the location, they are more likely to give correct explanations.

\begin{table}[H]
    \caption{\textcolor{black}{Spearman correlation coefficient matrix. Significant values (except diagonal) at $\alpha = 0.050$ are in bold.}}
    \centering
    \renewcommand{\arraystretch}{1.4} 
    \resizebox{\textwidth}{!}{%
    \begin{tabular}{clccccc}
        \toprule
        & & \multicolumn{2}{c}{\textcolor{black}{Accuracy}} & \multicolumn{3}{c}{\textcolor{black}{Experience}} \\
        \cmidrule(lr){3-4} \cmidrule(lr){5-7}
        & & \begin{tabular}[c]{@{}c@{}}\textcolor{black}{Avg Location}\\[-0.8ex] \textcolor{black}{Accuracy}\end{tabular} & \begin{tabular}[c]{@{}c@{}}\textcolor{black}{Ladybug Location}\\[-0.8ex] \textcolor{black}{Accuracy}\end{tabular} & \begin{tabular}[c]{@{}c@{}}\textcolor{black}{Months of}\\[-0.8ex] \textcolor{black}{Industry Experience}\end{tabular} & \begin{tabular}[c]{@{}c@{}}\textcolor{black}{Years}\\[-0.8ex] \textcolor{black}{Coding}\end{tabular} & \begin{tabular}[c]{@{}c@{}}\textcolor{black}{Self-Rated}\\[-0.8ex] \textcolor{black}{Expertise in C}\end{tabular} \\ \cline{3-7}
        
        \multirow{2}{*}{\rotatebox[origin=c]{90}{\textcolor{black}{Accuracy}}} 
        & \textcolor{black}{Avg Location Accuracy} & \textcolor{black}{1.000} & \textcolor{black}{0.288} & \textcolor{black}{-0.413} & \textcolor{black}{-0.323} & \textcolor{black}{0.221} \\
        & \textcolor{black}{Ladybug Location Accuracy} & \textcolor{black}{0.288} & \textcolor{black}{1.000} & \textcolor{black}{-0.296} & \textcolor{black}{-0.385} & \textcolor{black}{-0.274} \\
        \cline{3-7}
        
        \multirow{3}{*}{\rotatebox[origin=c]{90}{\textcolor{black}{Experience}}} 
        & \begin{tabular}[c]{@{}l@{}}\textcolor{black}{Months of Industry}\\[-0.8ex] \textcolor{black}{Experience}\end{tabular} & \textcolor{black}{-0.413} & \textcolor{black}{-0.296} & \textcolor{black}{1.000} & \textcolor{black}{\textbf{0.634}} & \textcolor{black}{0.190} \\
        & \textcolor{black}{Years Coding} & \textcolor{black}{-0.323} & \textcolor{black}{-0.385} & \textcolor{black}{\textbf{0.634}} & \textcolor{black}{1.000} & \textcolor{black}{0.225} \\
        & \begin{tabular}[c]{@{}l@{}}\textcolor{black}{Self-Rated}\\[-0.8ex] \textcolor{black}{Expertise in C}\end{tabular} & \textcolor{black}{0.221} & \textcolor{black}{-0.274} & \textcolor{black}{0.190} & \textcolor{black}{0.225} & \textcolor{black}{1.000} \\
        \bottomrule
    \end{tabular}
    }
    \label{tab:experience}
\end{table}
\textcolor{black}{Table ~\ref{tab:experience} shows the Spearman correlations among ``location'' accuracy scores and several measures of participant experience, with significant results in bold. We do not see evidence that months of industry experience, years coding in any language, or self-rated expertise in C correlates with average ``location'' accuracy score or the ``location'' accuracy score on bug \texttt{ladybug} (a sway bug). \texttt{Ladybug} (a sway bug) is shown here because most participants completed \texttt{ladybug}. One reason why these measures of experience are not correlated with ``location'' accuracy performance could be that even though undergraduate students have less industry experience and years coding than graduate students, they may have recently completed a course like Operating Systems where they practiced debugging C memory bugs.} 

\begin{table}[H]
    \centering
    \caption{Self-Reported Attention Level Popup Options}
    \label{tab:popup}
    \renewcommand{\arraystretch}{1.1}
    \begin{tabular}{p{7.2cm} l}
        \toprule
        \textbf{Option} & \textbf{Percent} \\
        \midrule
        I was focused on the task. & 91.03\% \\
        I was thinking about my performance on the task or how long it is taking. & 6.06\% \\
        I was thinking about things unrelated to the task. & 1.58\% \\
        I was distracted by sights or sounds in my environment. & 0.79\% \\
        My mind was blank. & 0.00\% \\
        \bottomrule
    \end{tabular}
\end{table}
Overall participants reported being focused on the tasks at hand.  In the attention question (prompted every five minutes), nine out of ten responses indicated attentiveness, as shown in Table~\ref{tab:popup}.  While we cannot know peoples' mental states with 100\% certainty, one possible explanation for the high level of reported attentiveness is the duration of the tasks (30 minutes per bug). Studies with longer tasks are more likely to lead to fatigue and inattentiveness. At a minimum, we interpret this finding to mean that we do not have evidence that participants were excessively distracted by the environment or other outside factors. We do not analyze the eye-tracking data in relation to the attentiveness survey due to the scarcity of data points showing inattentiveness.

\begin{table}[H]
\centering
\caption{Mean time required (in seconds) and mean scores per bug.}
\label{tab:bug_time_scores}

\begin{tabular}{l l r ccc ccc}
\toprule
\multicolumn{2}{c}{} & & \multicolumn{3}{c}{\textbf{\textcolor{black}{Cause}}} & \multicolumn{3}{c}{\textbf{\textcolor{black}{Location}}} \\
\cmidrule(lr){4-6} \cmidrule(lr){7-9}
Program & Codename & \textbf{\textcolor{black}{Time (s)}} & Confidence & Difficulty & Accuracy & Confidence & Difficulty & Accuracy \\
\midrule
sway    & ladybug    & 1330 & 3.80 & 2.48 & 3.20 & 3.50 & 2.95 & 3.20 \\
redis   & stonefly   & 1426 & 3.02 & 3.31 & 2.67 & 3.17 & 3.38 & 2.67 \\
sway    & hornet     & 1536 & 2.88 & 4.00 & 3.00 & 3.13 & 4.00 & 3.13 \\
redis   & mantis     & 1159 & 2.89 & 3.61 & 1.78 & 2.31 & 3.94 & 2.22 \\
openssl & firefly    & 1259 & 3.00 & 3.79 & 1.86 & 1.71 & 4.14 & 1.71 \\
redis   & silverfish & 1241 & 2.14 & 4.14 & 1.71 & 2.29 & 3.86 & 1.86 \\
sway    & spider     & 645  & 4.67 & 1.61 & 4.67 & 4.28 & 1.83 & 4.78 \\
openssl & weevil     & 933  & 2.39 & 3.67 & 3.67 & 3.11 & 3.28 & 3.44 \\
\bottomrule
\end{tabular}
\end{table}

\begin{table}[]
\setlength{\tabcolsep}{4.5pt}
\centering
\normalsize
\caption{Gaze metrics on code for different bugs. Mean number of fixations on code for all bugs is 838 over 1191 seconds, or an average of 42 per minute. \textcolor{black}{The column ``Program'' refers to the github project/repository that the code comes from. The column ``Bug'' contains the codename we assigned each task. For ``Fixation Count,'' ``Regression Rate,'' ``Duration (ms),'' ``Lines,'' and ``Functions,'' to calculate the mean and median values, we aggregate the statistics over all the participants who completed the task in that row. ``Fixation Count'' is the total number of fixations that occur during a task. A fixation is a stabilization of the eye on a stimulus for a period of time~\cite{sharafi2015eye}. ``Regression Rate'' is  the percentage of fixations where the token fixated on was fixated on more than once~\cite{Wallace2025Programmer}. ``Duration (ms)'' is the average fixation duration in milliseconds. ``Lines'' is the total number of unique lines that received fixations from the participant. ``Functions'' is the total number of unique functions that received fixations from the participant.}}
\label{tab:bug_metrics_formatted}

\begin{tabular}{l l rr rr rr rr rr}
\toprule
\textcolor{black}{Program} & 
Bug 
& \multicolumn{2}{c}{Fixation Count} 
& \multicolumn{2}{c}{Regression Rate} 
& \multicolumn{2}{c}{Duration (ms)} 
& \multicolumn{2}{c}{Lines} 
& \multicolumn{2}{c}{Functions} \\

\cmidrule(lr){3-4}
\cmidrule(lr){5-6}
\cmidrule(lr){7-8}
\cmidrule(lr){9-10}
\cmidrule(lr){11-12}

& & Mean & Median 
& Mean & Median 
& Mean & Median 
& Mean & Median 
& Mean & Median \\
\midrule

\textcolor{black}{sway}   &  ladybug    & 773 & 756   & 62 & 62 & 450 & 440 & 153 & 146 & 27 & 23 \\
\textcolor{black}{redis}   & stonefly   & 1024 & 1022 & 52 & 56 & 447 & 450 & 198 & 172 & 31 & 30 \\
\textcolor{black}{sway}    & hornet     & 1250 & 1560 & 51 & 54 & 412 & 426 & 239 & 215 & 25 & 12 \\
\textcolor{black}{redis}   & mantis     & 658 & 695   & 44 & 45 & 474 & 433 & 182 & 180 & 17 & 18 \\
\textcolor{black}{openssl} & firefly    & 890 & 937   & 44 & 42 & 403 & 416 & 260 & 302 & 43 & 43 \\
\textcolor{black}{redis}   & silverfish & 588 & 591   & 43 & 47 & 380 & 393 & 170 & 135 & 21 & 12 \\
\textcolor{black}{sway}    & spider     & 460 & 425   & 55 & 56 & 423 & 416 & 103 & 102 & 11 & 12 \\
\textcolor{black}{openssl} & weevil     & 1062 & 770  & 67 & 67 & 435 & 422 & 125 & 95 & 8 & 6 \\
\bottomrule
\end{tabular}
\end{table}

\begin{table}[h]
    
\centering
\renewcommand{\arraystretch}{1.2}

\centering
\caption{Descriptive statistics of the number of most fixated on \textbf{functions} that capture 75\% of all fixations on \textbf{functions}. \textcolor{black}{For the Mean, Min, Max, and SD (Standard Deviation), these statistics are aggregated over all the participants who completed the task. ``Functions in Top 75\%'' refers to the number of functions that contain 75\% of the participants' fixations. ``Percent in Top 75\%'' computes the percentage of functions in ``Functions in Top 75\%'' relative to the total number of functions fixated on. 
}}
\label{tab:bug_metrics_grouped_func}

\begin{tabular}{l l *{4}{p{6mm}} *{4}{p{6mm}}}
\toprule
\textcolor{black}{Program} & Bug & \multicolumn{4}{c}{Functions in Top 75\%} & \multicolumn{4}{@{}c@{}}{Percent in Top 75\%}\\
\cmidrule(lr){3-6} \cmidrule(lr){7-10}
& & Mean & Min & Max & SD & Mean & Min & Max & SD \\
\midrule
\textcolor{black}{sway}   &  ladybug    & 5.4 & 1 & 18 & 4.3  & 20\% & 8\% & 38\% & 7\% \\
\textcolor{black}{redis}   & stonefly   & 4.2 & 2 & 11 & 2.2  & 18\% & 7\% & 67\% & 14\% \\
\textcolor{black}{sway}    & hornet     & 8.5 & 2 & 46 & 15.2 & 34\% & 15\% & 60\% & 14\% \\
\textcolor{black}{redis}   & mantis     & 3.3 & 1 & 7  & 1.8  & 20\% & 11\% & 33\% & 6\% \\
\textcolor{black}{openssl} & firefly    & 6.4 & 1 & 12 & 4.3  & 17\% & 9\% & 33\% & 8\% \\
\textcolor{black}{redis}   & silverfish & 4.7 & 2 & 13 & 4.3  & 31\% & 10\% & 67\% & 18\% \\
\textcolor{black}{sway}    & spider     & 2.2 & 1 & 4  & 0.8  & 24\% & 15\% & 40\% & 9\% \\
\textcolor{black}{openssl} & weevil     & 2.8 & 1 & 4  & 0.8  & 51\% & 16\% & 100\% & 26\% \\
& all bugs   & 4.6 & 1 & 46 & 5.4  & 25\% & 7\% & 100\% & 17\% \\
\bottomrule
\end{tabular}
\end{table}
\begin{figure}[h]
\centering
\begin{subfigure}{0.8\textwidth}
    \centering
    \includegraphics[width=\linewidth]{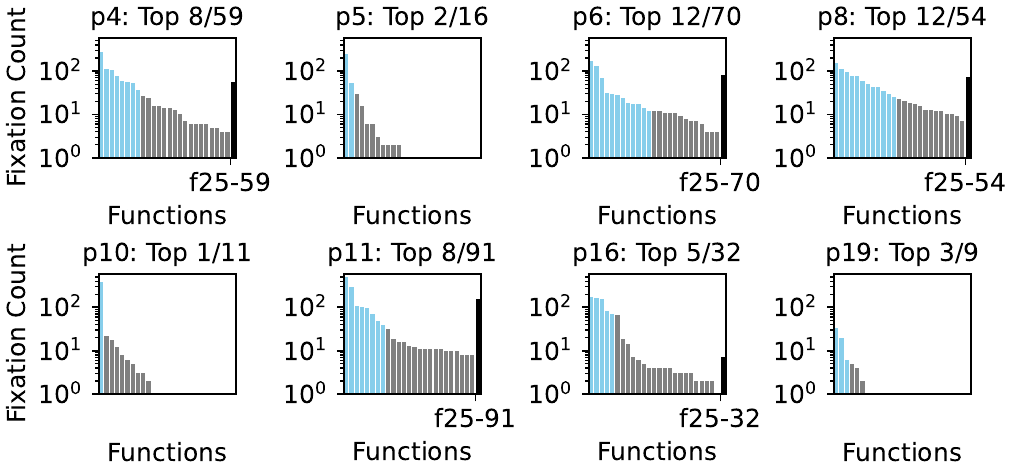}
    \caption{Bar charts showing the distribution of fixations over \textbf{functions} read by participants locating the bug \texttt{firefly} \textcolor{black}{(an \texttt{openssl} bug)}.  Blue indicates the top \textbf{functions} that capture 75\% of the total fixations on \textbf{functions}. Reading effort is concentrated on a few functions.}
    \Description{This figure shows 8 bar charts. Each chart's y-axis is "Fixation Count." The x-axis shows the number of functions read. The height of each bar is the number of fixations on the function. Some of the bars are blue, and some are grey. The blue bars indicate which functions make up the top 75\% of fixations on functions. This figure's purpose is to illustrate that most reading effort (fixations) is concentrated on a small number of functions.}
    \label{fig:fix_quartiles_func}
    \vspace{2em}
\end{subfigure}

\begin{subfigure}{0.8\textwidth}
    \centering
    \includegraphics[width=\linewidth]{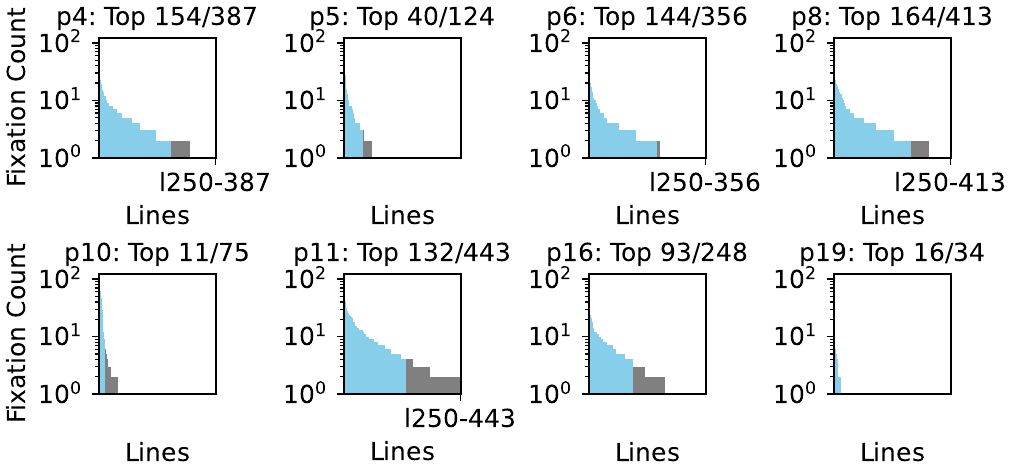}
    \caption{Bar charts showing the distribution of fixations over \textbf{lines} read by participants locating the bug \texttt{firefly} \textcolor{black}{(an \texttt{openssl} bug)}.  Blue indicates the top \textbf{lines} that capture 75\% of the total fixations on \textbf{lines}.  Reading effort is distributed more widely than for functions.}
    \Description{This figure shows 8 bar charts. Each chart's y-axis is "Fixation Count." The x-axis shows the number of lines read. The height of each bar is the number of fixations on the line. Some of the bars are blue, and some are grey. The blue bars indicate which lines make up the top 75\% of fixations on lines of code. This figure's purpose is to illustrate that most reading effort (fixations) is concentrated on a small number of lines of code.}
    \label{fig:fix_quartiles_line}
\end{subfigure}
\caption{Bar charts showing the distribution of fixations over functions and lines of code.}
\Description{}
\end{figure}
\begin{table}[h]
\renewcommand{\arraystretch}{1.2}
\centering
\caption{Descriptive statistics of the number of most fixated on \textbf{lines} that capture 75\% of all fixations on \textbf{lines} of code. \textcolor{black}{For the Mean, Min, Max, and SD (Standard Deviation), these statistics are aggregated over all the participants who completed the task. ``Lines in Top 75\%'' refers to the number of lines that contain 75\% of the participants' fixations. ``Percent in Top 75\%'' computes the percentage of lines in ``Lines in Top 75\%'' relative to the total number of lines fixated on.}}
\label{tab:bug_metrics_grouped_line}

\begin{tabular}{l l *{4}{p{6mm}} *{4}{p{6mm}}}
\toprule
\textcolor{black}{Program} & Bug & \multicolumn{4}{c}{Lines in Top 75\%} & \multicolumn{4}{c}{Percent in Top 75\%} \\
\cmidrule(lr){3-6} \cmidrule(lr){7-10}
& & Mean & Min & Max & SD & Mean & Min & Max & SD \\
\midrule
\textcolor{black}{sway}   &  ladybug    & 43.6 & 14 & 141 & 34.7 & 28\% & 17\% & 62\% & 10\% \\
\textcolor{black}{redis}   & stonefly   & 51.6 & 12 & 129 & 29.3 & 28\% & 16\% & 68\% & 12\% \\
\textcolor{black}{sway}    & hornet     & 91.9 & 44 & 207 & 54.9 & 38\% & 29\% & 62\% & 11\% \\
\textcolor{black}{redis}   & mantis     & 61.5 & 19 & 123 & 29.5 & 34\% & 22\% & 44\% & 7\% \\
\textcolor{black}{openssl} & firefly    & 94.3 & 11 & 164 & 63.6 & 35\% & 15\% & 47\% & 10\% \\
\textcolor{black}{redis}   & silverfish & 63.1 & 11 & 168 & 61.0 & 35\% & 10\% & 53\% & 14\% \\
\textcolor{black}{sway}    & spider     & 33.6 & 10 & 61  & 17.8 & 32\% & 25\% & 36\% & 4\% \\
\textcolor{black}{openssl} & weevil     & 36.3 & 23 & 60  & 11.8 & 31\% & 20\% & 42\% & 6\% \\
& all bugs   & 56.0 & 10 & 207 & 41.7 & 31\% & 10\% & 68\% & 10\% \\
\bottomrule
\end{tabular}
\end{table}

\subsection{RQ2: Visual Attention Patterns}
\label{sec:rq2}
\begin{figure}[hb]
    \centering
    \includegraphics[width=0.7\linewidth]{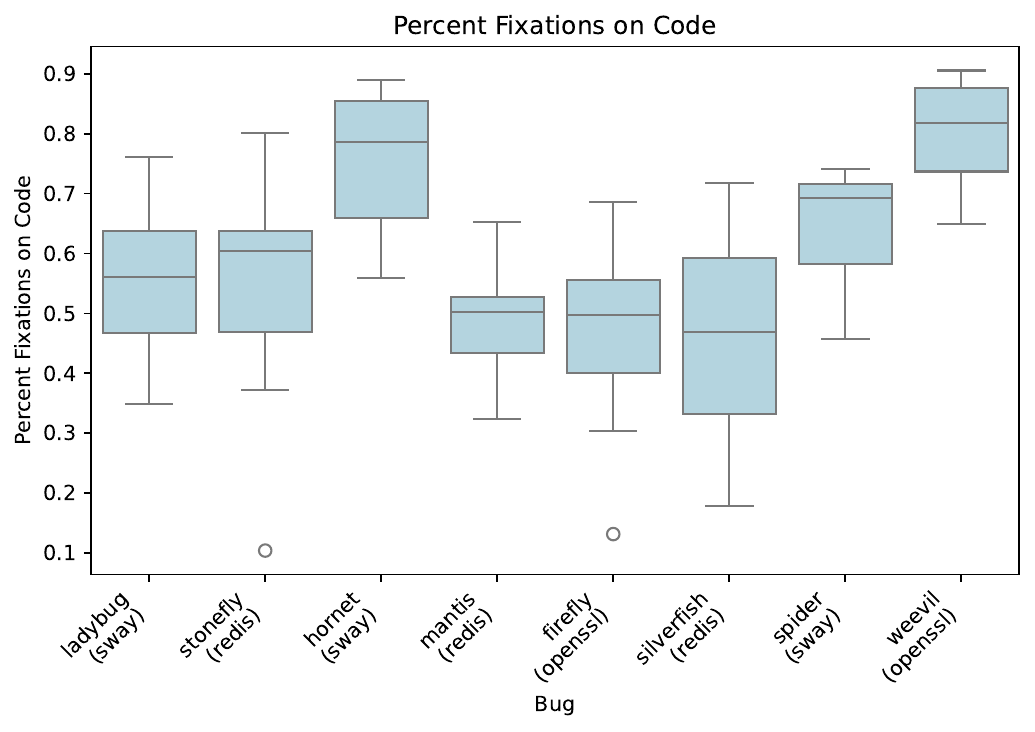}
    \caption{Percent fixations on code for each bug: time looking at code versus other regions while working on a particular bug.  Code corresponds to area B in Figure~\ref{fig:interface}.  Note not all participants worked on all bugs.}
    \Description{This figure shows a box plot. The x-axis has bug codenames. The y-axis has percentages from 0 to 100. This is a box and whiskers plot.}
    \label{fig:codefix}
\end{figure}

We summarize broad patterns of visual attention in Tables~\ref{tab:bug_metrics_formatted}~-~\ref{tab:bug_metrics_grouped_line} and Figures~\ref{fig:fix_quartiles_func}~-~\ref{fig:codefix}.  Table~\ref{tab:bug_metrics_formatted} shows a considerable spread in fixation count among the bugs. The bug \texttt{spider} \textcolor{black}{(a \texttt{sway} bug)} has the lowest fixation count, at 460.   
The most likely explanation is that this bug is also the easiest, as Table~\ref{tab:bug_time_scores} shows \texttt{spider} \textcolor{black}{(a \texttt{sway} bug)} with the highest confidence, lowest difficulty, and highest accuracy.  Unsurprisingly, the bug with the highest mean fixation count (\texttt{hornet} \textcolor{black}{(a \texttt{sway} bug)}) was among the highest difficulty.  However, fixation count is not a perfect indicator of difficulty. For example, \texttt{silverfish} \textcolor{black}{(a \texttt{redis} bug)} has a relatively low fixation count, but a high difficulty. This may be explained when looking at the average time for \texttt{silverfish} in Table~\ref{tab:bug_time_scores}. This high time but low fixation count on code indicates that participants spent a lot of time reading the bug report. The \texttt{silverfish} \textcolor{black}{(a \texttt{redis} bug)} bug report was not particularly long, but participants may have had difficulty interpreting the report. In all cases, the mean and median fixation counts and regression rates per bug are relatively similar, indicating that outliers are not markedly affecting the mean. 
In this work, ``regression rate'' is the percentage of fixations where the token fixated on was fixated on more than once~\cite{Wallace2025Programmer}. The mean fixation duration in milliseconds was stable throughout, with no clear association with bug difficulty.


A novel observation of this paper is that people read several functions outside of the one where the bug is located (a mean of 22), though visual attention is concentrated among relatively few. 
\begin{highlightbox}
\textbf{\finding{\label{finding:75pct}Table~\ref{tab:bug_metrics_grouped_func} shows that programmers spend 75\% of their visual effort on 25\% of the functions that they view.}}
\end{highlightbox}

  In rough practical terms, for a typical bug a typical person read 23 functions but spent 75\% of their visual effort on 5--6 of them.  Figure~\ref{fig:fix_quartiles_func} shows a representative example of this pattern with the bug \texttt{firefly} \textcolor{black}{(an \texttt{openssl} bug)}:  The x-axis shows one bar for each function viewed and the height of the bar on the y-axis is the number of fixations.  Eight study participants did this task.  And while the number of functions viewed varied substantially (min 9, max 91), the pattern is consistent of concentrated visual attention.  Put succinctly, around 75\% of visual attention is paid to 25\% of functions that are read.

We observe that the distribution of visual effort over lines is concentrated to a lesser degree than over functions.  Consider Table~\ref{tab:bug_metrics_grouped_line}.  Around 31\% of the lines read occupied 75\% of the total number of fixations on code lines. The example in Figure~\ref{fig:fix_quartiles_line} for bug \texttt{firefly} \textcolor{black}{(an \texttt{openssl} bug)} shows this pattern.  While some participants focused on a small number of lines (e.g., p10), it was common to spread out one's visual attention more, as indicated with the longer sections marked in Figures~\ref{fig:fix_quartiles_func} and ~\ref{fig:fix_quartiles_line} in blue.

Around 58\% of the participants' fixations were spent on code on average, though this number was highly variable.  The remaining fixations mostly spent reading the bug report details (a near-trivial amount of time was spent navigating the IDE).  Recall from Figure~\ref{fig:interface} that participants had code, report, navigation, and AI areas visible and that they were not permitted to leave the IDE.  Figure~\ref{fig:codefix} shows the percent of visual attention (measured by fixations) spent on code during each bug.  We caution against drawing strong conclusions generalizing participant behavior in terms of region fixated on because we observe a range from 10\% to 90\% for some bugs. Some participants spend the vast majority of their effort reading code, while others spend that effort reading the bug report.
\begin{table*}[h]
\centering
\caption{Participant AI Query Counts and Percentage of Gazes on Each Screen Region. Code and Report are combined because participants often opened code and reports in the same region. See Figure \ref{fig:codefix} for a breakdown between code and report fixations. ``Other'' includes the iTrace window, out of bounds gazes, and gazes without coordinates. ``UNK'' indicates where participants changed what was displayed in the region frequently. For example, participant 4 frequently flipped back and forth between the file explorer and code in the same screen region.}
\label{tab:ai_queries}
\vspace{-2mm}
\begin{tabular}{cccccc}
\toprule
Partic. & \begin{tabular}[c]{@{}c@{}}Queries\\to AI\end{tabular} & \begin{tabular}[c]{@{}c@{}}Percent Gazes at\\ AI Window (D)\end{tabular} & \begin{tabular}[c]{@{}c@{}}Percent Gazes at\\ Code/Report (A/B)\end{tabular} & \begin{tabular}[c]{@{}c@{}}Percent Gazes at\\ File Explorer (C)\end{tabular} & \begin{tabular}[c]{@{}c@{}}Percent Gazes at\\ ``Other''\end{tabular} \\
\midrule
p1 & 0 & 0.00\% & 74.60\% & 14.40\% & 11.00\% \\
p2 & 5 & 7.89\% & 78.55\% & 6.74\% & 6.83\% \\
p3 & 0 & 0.00\% & 76.43\% & 9.69\% & 13.88\% \\
p4 & 6 & 4.09\% & UNK     & UNK & 9.58\% \\
p5 & 8 & 7.46\% & 73.62\% & 10.94\% & 7.99\% \\
p6 & 0 & 0.18\% & 78.99\% & 15.71\% & 5.12\% \\
p7 & 0 & 0.05\% & 39.89\% & 2.55\% & 57.51\% \\
p8 & 2 & 1.85\% & 84.24\% & 3.45\% & 10.47\% \\
p9 & 4 & 2.09\% & 76.35\% & 10.47\% & 11.09\% \\
p10 & 2 & 1.44\% & 78.82\% & 6.38\% & 13.36\% \\
p11 & 0 & 0.00\% & UNK   & UNK & 6.06\% \\
p12 & 2 & 2.72\% & UNK   & UNK & 35.92\% \\
p13 & 11 & 12.42\% & 78.41\% & 5.08\% & 4.10\% \\
p14 & 3 & 3.97\% & UNK & UNK & 18.41\% \\
p15 & 8 & 14.51\% & UNK & UNK & 8.33\% \\
p16 & 4 & UNK & UNK & 4.41\% & 15.08\% \\
p17 & 6 & 3.33\% & 85.94\% & 6.30\% & 4.43\% \\
p18 & 7 & 1.01\% & UNK & UNK & 46.83\% \\
p19 & 4 & 2.37\% & 73.43\% & 7.46\% & 16.75\% \\
p20 & 0 & 0.39\% & UNK & UNK & 6.07\% \\
p21 & 2 & 1.64\% & 80.34\% & 11.59\% & 6.43\% \\
\bottomrule
\end{tabular}
\vspace{-5mm}
\end{table*}

A final observation is that relatively little outside help was requested, as measured by the number of queries people asked in the AI help window and visual effort spent on this area. Table~\ref{tab:ai_queries} shows only a handful of queries were made per participant, with 6 of 21 participants not requesting AI help. 
Using the AI was not correlated with confidence, accuracy, or difficulty.  
As instructed, participants did not ask the AI to solve/find the bug or copy and paste significant snippets of code. Instead they asked questions related to how to use \texttt{calloc} (P4, P5, P9, P13, P14, P15, P17, P18, P19) and what is \texttt{zmalloc} (P4, P5, P13, P18, P21), for example.

\begin{figure}[h]
    \centering
    \includegraphics[width=0.8\linewidth]{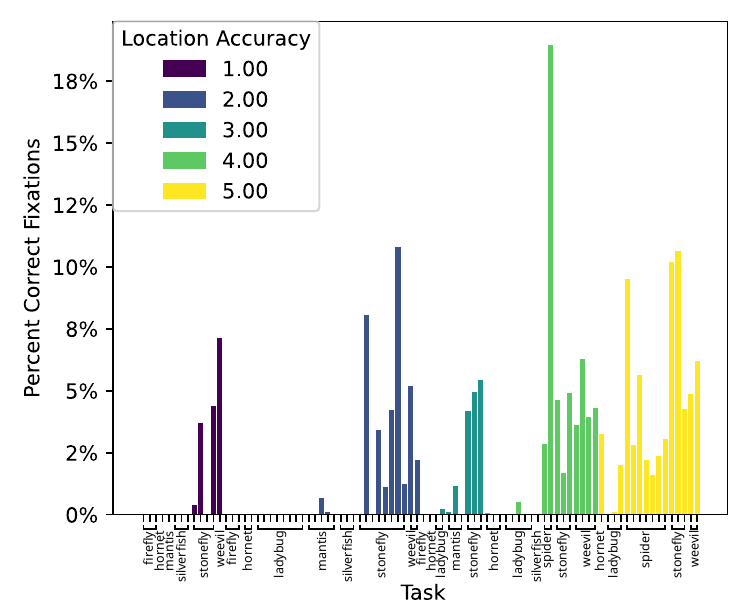}
    \caption{Percent fixations on a ``correct bug \textcolor{black}{location}'' from lowest to highest ``\textcolor{black}{location accuracy}'' score for each participant-task.  More fixations on correct locations are observed with higher accuracy, but correct lines were still read even at low accuracy.}
    \Description{This bar chart has Percent Correct Fixations on the y-axis. Each bar represents one task. The color of the bar indicates the \textcolor{black}{Location Accuracy} score earned for that task. As the \textcolor{black}{Location Accuracy} increases, there are more bars that have correct fixation percentages above 0.}
    \label{fig:fix_on_correct}
\end{figure}
\subsection{RQ3: Attention and Task Outcome}

Overall we observed differences in terms of regression rate and mean euclidean distance between fixations for higher versus lower performance on tasks.  \textcolor{black}{Euclidean distance is the length of a line segment between 2 points. The mean euclidean distance between fixations refers to the mean distance from the x,y coordinates of each fixation to the next fixation. Regression rate is defined in Section~\ref{sec:rq2}.} Although we did not find statistically significant differences in fixation count, fixation duration, or the number of unique lines or methods read, a lack of statistical significance does not imply the absence of an underlying effect; it only means that the current data do not provide sufficient evidence to detect one. Our findings are conceptually similar to reports from Software Engineering and other research areas that higher or lower performance on tasks can be associated with greater focus on certain thematically-relevant components~\cite{carmichael2010does, eivazi2011predicting, fritz2014using, jin2021study, steichen2014inferring}.  This section discusses the nature of these differences in our study and key caveats.

First consider Figure~\ref{fig:fix_on_correct} which shows the percent of fixations on a ``correct bug \textcolor{black}{location}.''  A ``correct bug \textcolor{black}{location}'' is a line in the code that would have been marked as a correct answer to the ``\textcolor{black}{location accuracy}'' question (see Section~\ref{sec:dataprocessing} for definitions). Each bar in Figure~\ref{fig:fix_on_correct} is a task by a participant.  The bar height is the percent of fixations during that task that the participant looked at a correct bug \textcolor{black}{location}.   

\begin{highlightbox}
\textbf{\finding{\label{finding:fail}While the percentages \textcolor{black}{in Figure~\ref{fig:fix_on_correct}} themselves should not be over-interpreted, what we observe is that even during tasks with a low ``\textcolor{black}{location accuracy},'' participants often fixated on correct bug lines at least briefly.  This means that participants failed to mark these lines as bug locations even though we have evidence that they read those locations.}}
\end{highlightbox}

\begin{figure}[h]
\includegraphics[width=0.6\linewidth]{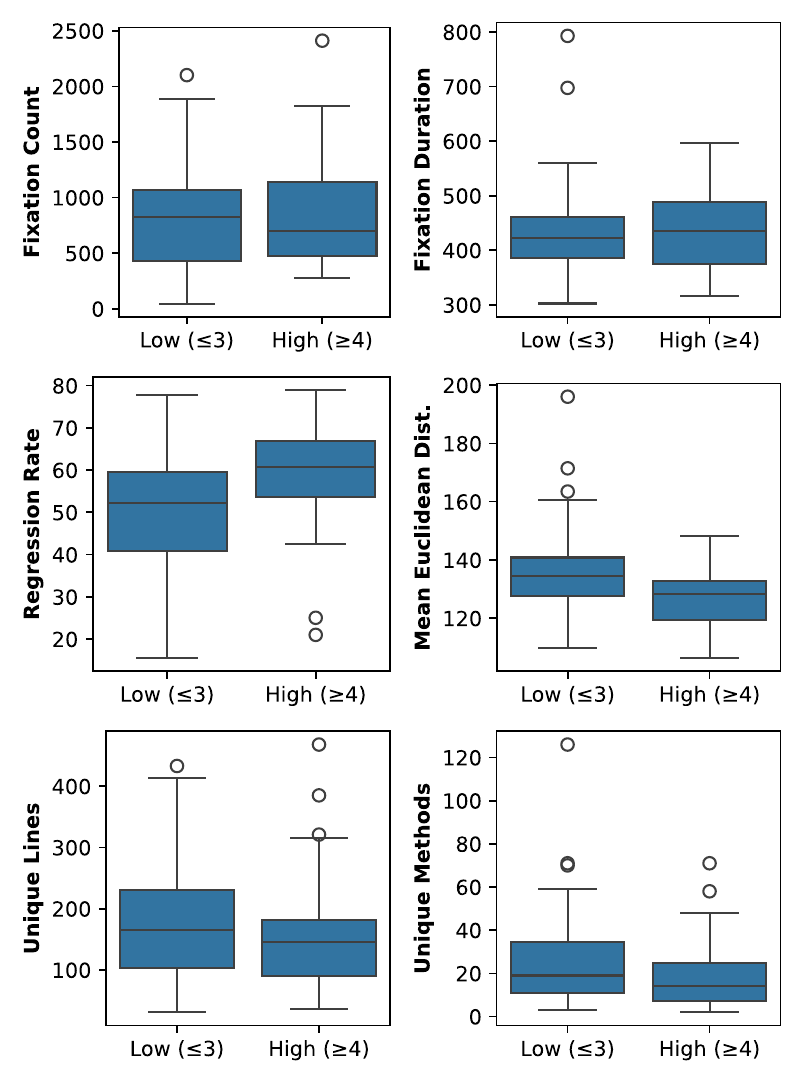}
\caption{Boxplots to accompany statistical summary of eye metrics in Table ~\ref{tab:rq3metrics}. Low group means accuracy scores $\leq$3.  High group means scores $\geq$4.}
\Description{This figure is made up of 6 smaller figures. Each of these smaller figures is a box plot comparing a metric for the low group and the high group. The six metrics are: fixation count, fixation duration, regression rate, mean euclidean distance, unique lines, unique methods.}
\label{fig:boxplots}
\vspace{-3mm}
\end{figure}

\begin{figure}[h]
    \centering
    \includegraphics[width=0.7\linewidth]{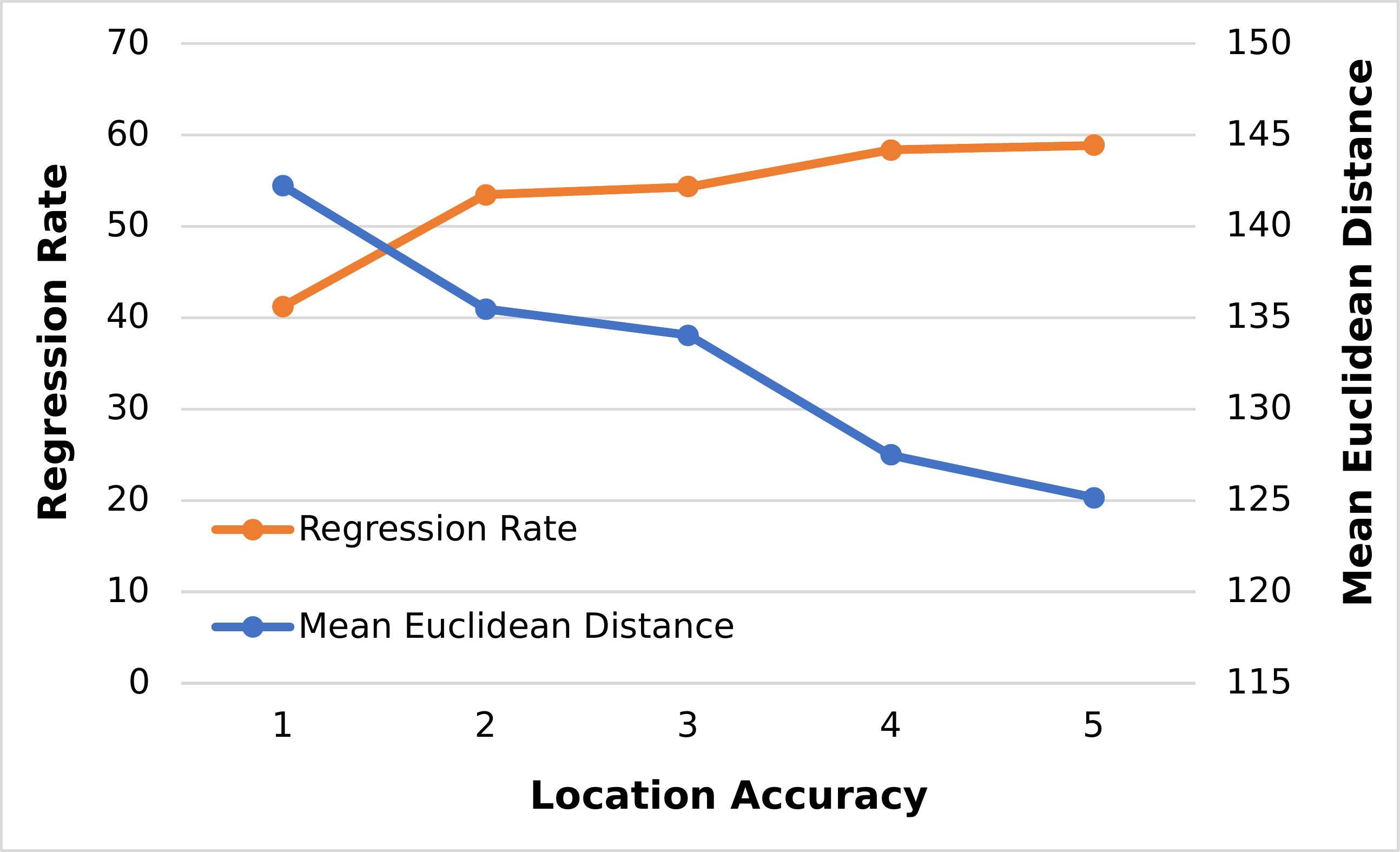}
    \caption{As ``\textcolor{black}{Location Accuracy}'' increases, the mean distance between fixations tends to decrease, and the regression rate increases.}
    \Description{This figures shows a line plot. There are two y-axes on this chart. On the left side is the Regression Rate from 0 to 70. On the right side is the Mean Euclidean Distance from 115 to 150. The x-axis is the \textcolor{black}{Location Accuracy}. The Regression rate line increases from 1 to 2 to 3 to 4 to 5. The Mean Euclidean distance line decreases as the \textcolor{black}{Location Accuracy} increases.}
    \label{fig:eucl_reg_chart}
\end{figure}

\begin{table}[h]
\centering
\caption{Statistical summary of eye metrics and accompanying boxplots.  
Low group means accuracy scores $\le 3$.  
High group means accuracy scores $\ge 4$.\textcolor{black}{The \textit{p} value in the last column is the result of a Mann-Whitney U test.}}
\label{tab:rq3metrics}
\setlength{\tabcolsep}{3.6pt}
\begin{tabular}{l l r r r r r}
\toprule
Metric & Group & Mean & Min & Max & SD & p \\
\midrule
\multirow{2}{*}{fixation count}       
    & Low  & 820.4 & 45.0   & 2101.0 & 493.5 & \multirow{2}{*}{0.392} \\
    & High & 888.1 & 273.0  & 2411.0 & 540.2 &  \\
\midrule

\multirow{2}{*}{fixation duration}    
    & Low  & 437.1 & 302.4  & 792.8  & 84.8  & \multirow{2}{*}{0.495} \\
    & High & 432.3 & 316.2  & 596.4  & 67.1  &  \\
\midrule

\multirow{2}{*}{regression rate}      
    & Low  & 50.7  & 15.6   & 77.7   & 14.6  & \multirow{2}{*}{0.003} \\
    & High & 58.6  & 21.0   & 78.9   & 12.7  &  \\
\midrule

\multirow{2}{*}{mean euclidean dist.} 
    & Low  & 136.8 & 109.9  & 196.0  & 15.1  & \multirow{2}{*}{0.0004} \\
    & High & 126.4 & 106.3  & 148.2  & 9.8  &  \\
\midrule

\multirow{2}{*}{unique lines read}    
    & Low  & 182.5 & 31.0   & 433.0  & 108.8 & \multirow{2}{*}{0.184} \\
    & High & 162.0 & 36.0   & 468.0  & 99.0  &  \\
\midrule

\multirow{2}{*}{unique methods read}  
    & Low  & 26.1  & 3.0    & 126.0  & 22.4  & \multirow{2}{*}{0.068} \\
    & High & 19.2  & 2.0    & 71.0   & 15.8  &  \\
\bottomrule
\end{tabular}
\end{table}

As an initial point for understanding differences in attention based on task performance, we calculated six metrics of visual attention and split them based on high and low performance.  We define ``high'' as ``\textcolor{black}{location accuracy}'' scores 4--5 and ``low'' as ``\textcolor{black}{location accuracy}'' scores 1--3. We chose to group the scores in this way for three reasons. First, for the tasks where participants scored a 3, we looked at the rest of their tasks and determined that most participants who scored a 3 on one task also had more lower scores (1s and 2s) than higher scores (4s and 5s). Second, looking at the regression rate and euclidean distance averages used to make Figure~\ref{fig:eucl_reg_chart}, the averages for the ``\textcolor{black}{location accuracy}'' score of 3 was much closer to the averages for the ``\textcolor{black}{location accuracy}'' score of 2 than to the averages for the ``\textcolor{black}{location accuracy}'' score of 4, indicating that the visual attention patterns of 3s are closer to the visual attention patterns of 2s than 4s. Third, in order to receive a score of 3 for ``\textcolor{black}{location accuracy,}'' participants needed to have a ``limited idea where to look'' according to the rubric, and we do not deem that to be classified as ``high performance.'' Notably, if we choose to group the scores in different ways, (1s and 2s) versus (3s, 4s, and 5s) or (1s and 2s) versus (4s and 5s), the key findings and statically significant results do not change.

To preserve a conservative analysis, we use five metrics from earlier in this paper: fixation count and duration, regression rate, unique lines and methods read.  One additional metric is Mean Euclidean Distance, the mean distance from the x,y coordinates of each fixation to the next fixation.  Higher mean distance indicates more eye movement range.  We performed a Mann-Whitney U test of low/high scores for each metric. We chose this test to be conservative in not assuming a normal distribution and to be robust to outliers.
Table~\ref{tab:rq3metrics} and Figure~\ref{fig:boxplots} show comparisons for each metric.   
\begin{highlightbox}
\textbf{\finding{\label{finding:statsig}We observe a statistically-significant difference in regression rate and mean euclidean distance between bug localizations with low ``\textcolor{black}{location accuracy}'' and high ``\textcolor{black}{location accuracy}''.}} 
\end{highlightbox}

\edits{We discuss the implications of this key finding in Section~\ref{subsec:implications}.}
We observe differences in other metrics but these differences are not statistically-significant\edits{, so we cannot make claims about these metrics in our study}.  Figure~\ref{fig:eucl_reg_chart} shows regression rate and mean euclidean distance averaged for each ``\textcolor{black}{location accuracy}'' score.  The general pattern is that regression rate is higher as ``\textcolor{black}{location accuracy}'' increases, while mean euclidean distance is lower.

\begin{figure}[H]
\centering
\begin{subfigure}{0.9\textwidth}
    \centering
    \includegraphics[width=0.9\linewidth]{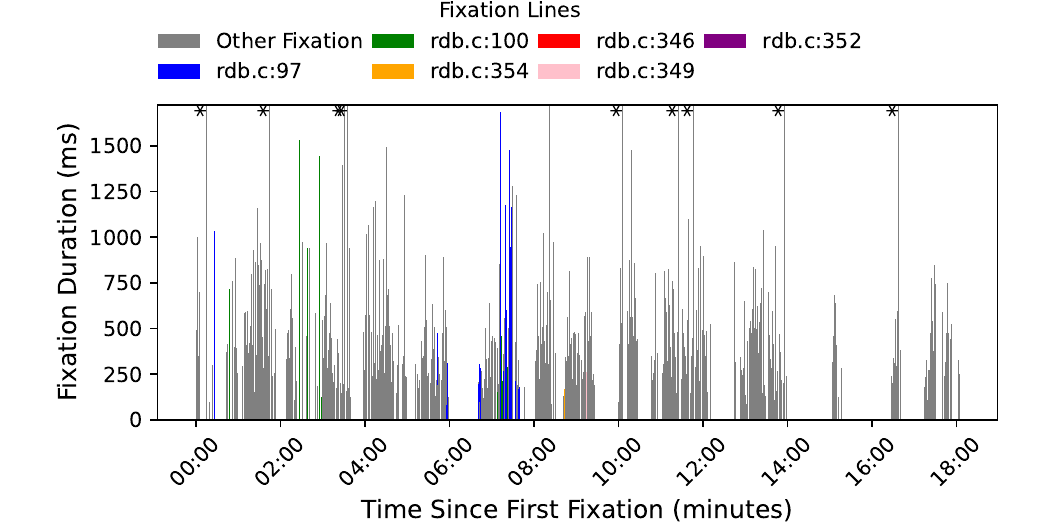}
    \caption{Participant p7, success at bug \texttt{stonefly} \textcolor{black}{(a redis bug)}.}
    \label{fig:stonefly_success}
\end{subfigure}

\begin{subfigure}{0.9\textwidth}
    \centering
    \includegraphics[width=0.9\linewidth]{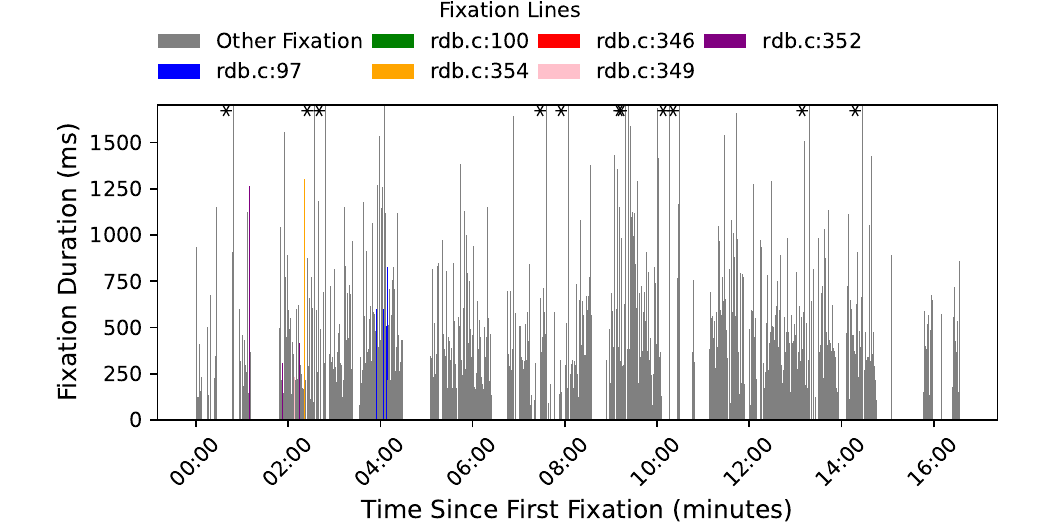}
    \caption{Participant p3, failure at bug  \texttt{stonefly} \textcolor{black}{(a redis bug)}.}
    \label{fig:stonefly_failure}
\end{subfigure}
\caption{Comparing successful and failure for cases for bug \texttt{stonefly} \textcolor{black}{(a \texttt{redis} bug)}. \textcolor{black}{Asterisks indicate truncated fixation durations to improve readability.}}
\end{figure}
\begin{figure}[]\ContinuedFloat
\centering
\begin{subfigure}{1.0\textwidth}
    \centering
    \includegraphics[width=\linewidth]{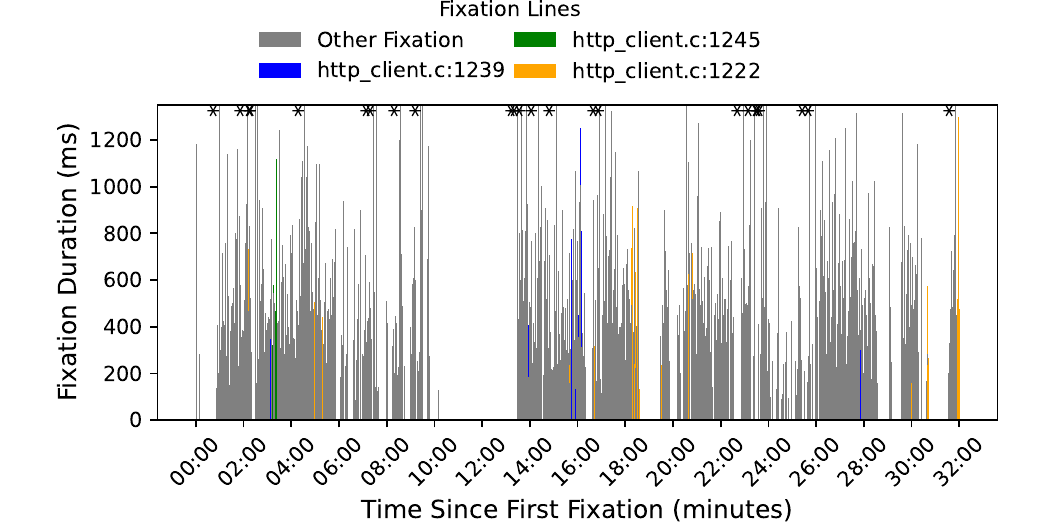}
    \caption{Participant p11, success at bug \texttt{weevil} \textcolor{black}{(an openssl bug)}.}
    \label{fig:weevil_success}
\end{subfigure}
\hfill
\begin{subfigure}{1.0\textwidth}
    \centering
    \includegraphics[width=\linewidth]{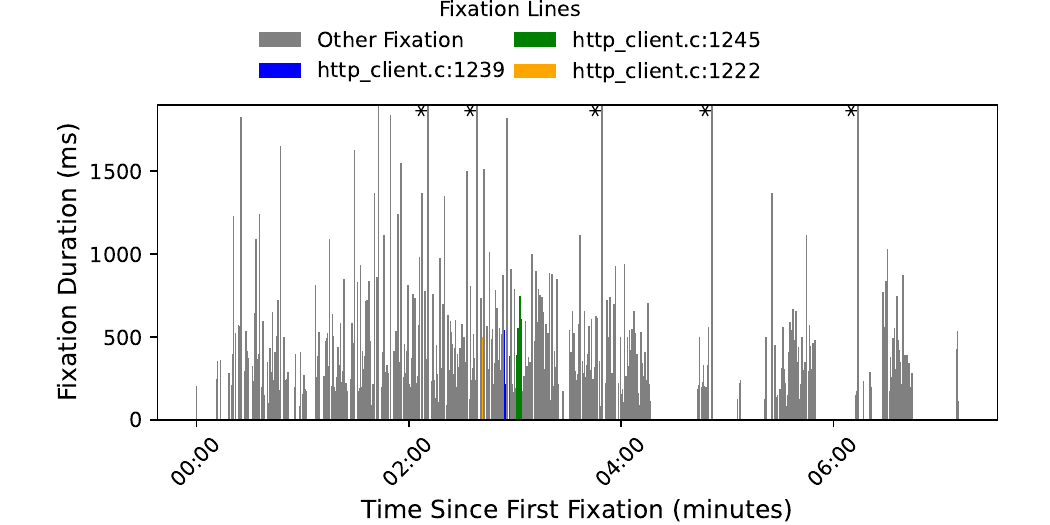}
    \caption{Participant p5, failure at bug \texttt{weevil} \textcolor{black}{(an openssl bug)}.}
    \label{fig:weevil_failure}
\end{subfigure}
\caption{Comparing successful and failure for cases for bug \texttt{weevil} \textcolor{black}{(an openssl bug)}. \textcolor{black}{Asterisks indicate truncated fixation durations to improve readability.}}
\Description{This figure has 4 subfigures. The subfigures each show a timeline on the x-axis. The y-axis is labeled as "Fixation Duration." Along the timeline, there are bars of different heights representing fixations of different durations. Most of the fixation bars are grey. Some of the fixation bars are other colors. These bars indicate fixations on a "correct line." There are more of these colored bars for the success case than the failure case.}
\end{figure}


Consider the examples in Figures~\ref{fig:stonefly_success}--\ref{fig:weevil_failure} to illustrate potential differences in success versus failure cases.  In these figures, each bar is a fixation on code (specifically area B indicated in Figure~\ref{fig:interface}). Asterisks indicate truncated fixation durations to improve readability. Areas with zero-height bars indicate fixations on other parts of the screen or offscreen.  The height of the bar is the duration of the fixation in milliseconds.  Colors indicate ``correct bug lines'' (recall that only one correct bug line is necessary even if multiple are possible).  Figure~\ref{fig:stonefly_success} shows a successful case on bug \texttt{stonefly} \textcolor{black}{(a \texttt{redis} bug)} while Figure~\ref{fig:stonefly_failure} shows a failure case.  Likewise Figure~\ref{fig:weevil_success} shows a successful case for bug \texttt{weevil} \textcolor{black}{(an \texttt{openssl} bug)} while Figure~\ref{fig:weevil_failure} shows a failure case.

For bug \texttt{stonefly} \textcolor{black}{(a \texttt{redis} bug)}, Figure~\ref{fig:stonefly_success} shows the participant reading a correct bug location at line 97 of \texttt{rdb.c} around the seven minute time mark.  The participant takes another eleven minutes to read other locations and conclude the session and write a correct answer to the problem.  In contrast, Figure~\ref{fig:stonefly_failure} shows a different participant reading several areas of code including correct bug lines, but for fewer fixations and less time per fixation -- ultimately that participant provided an incorrect answer.  A similar pattern is evident in Figure~\ref{fig:weevil_success} where a participant had several fixations on line 1222 of \texttt{http\_client.c} around the 18 minute mark, then continued to read other areas of code, before regressing to that location again and writing it as an answer at the end of the session.  And the failure case in Figure~\ref{fig:weevil_failure} again shows fixations on the correct location early in the session but for less time, with the participant ultimately providing an incorrect answer.
\begin{highlightbox}
\textbf{\finding{\label{finding:timeline}In success cases, participants often spend more time on the buggy line or regress back to the buggy line several times while in failure cases, participants often see the buggy line, but do not register it as the answer.}}
\end{highlightbox}

\begin{figure}[h]
    \centering
    \includegraphics[width=\linewidth]{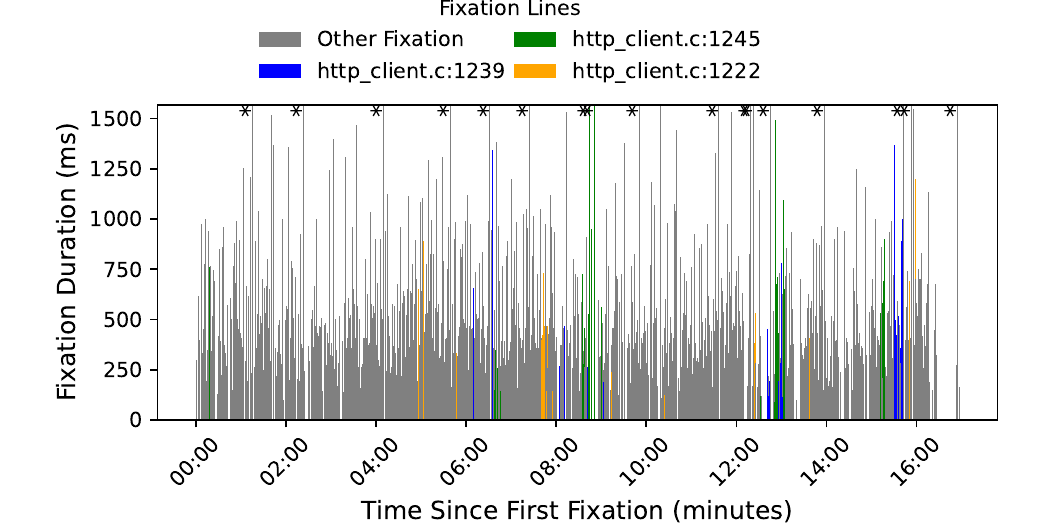}
    \caption{\textcolor{black}{Participant p17, failure at bug \texttt{weevil} (an \texttt{openssl} bug). Asterisks indicate truncated fixation durations to improve readability.}}
    \Description{This figure has a timeline on the x-axis. The y-axis is labeled as "Fixation Duration." Along the timeline, there are bars of different heights representing fixations of different durations. Most of the fixation bars are grey. Some of the fixation bars are other colors. These bars indicate fixations on a "correct line." For Participant p17, failure at bug weevil, we see many fixations on the ``correct'' line.}
    \label{fig:weevil_p17}
    \vspace{-2mm}
\end{figure}

\textcolor{black}{Figure~\ref{fig:fix_on_correct} shows that some participants who failed to report the bug correctly still had a relatively high number of fixations on the buggy line. Figure~\ref{fig:weevil_p17} shows an example of this: Participant 17 failed to find the ``correct'' line for \texttt{weevil} (an \texttt{openssl} bug), but also had many fixations on ``correct'' locations. One reason for this may be that the \texttt{weevil} bug report included a reference to the function containing the buggy line. If the participant focuses on the correct function, then the participant's chance of fixating on the buggy line increases.} 

\begin{figure}[h]
    \centering
    \includegraphics[width=0.8\linewidth]{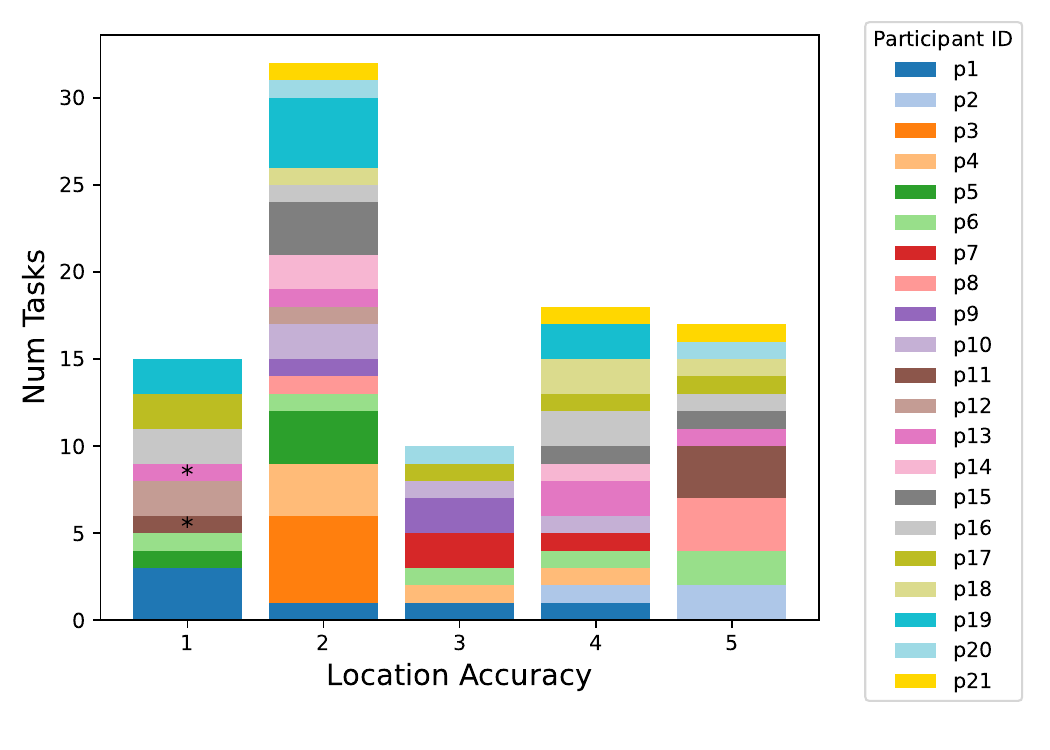}
    \caption{Distribution of bugs per participant grouped by ``\textcolor{black}{Location Accuracy}.''  Accuracy level 2 was most common, but participants were in all groups.  Asterisks \textcolor{black}{show that two scores were rounded from a score of 0 to a score of 1 for this visualization}.}
    \Description{This figure shows a stacked bar chart. The x-axis is "\textcolor{black}{Location Accuracy}." The y-axis is "Number of Tasks." Each bar has slices of different colors corresponding to a task done by a specific participant. For example, participant 1 is blue, and there are 3 blue slices in "\textcolor{black}{Location Accuracy}" = 1, meaning that participant 1 completed 3 tasks that earned a "\textcolor{black}{Location Accuracy}" score of 1.}
    \label{fig:part_task_dist}
\end{figure}

A caveat to these findings is summarized in Figure~\ref{fig:part_task_dist}: the low and high outcome groups are not balanced by participant because some participants performed less well overall, so some portion of our measurement could be of participant skill versus globally-relevant signs of success or failure. 

\begin{highlightbox}
\textbf{Thus we confine ourselves to conclusions that: 1) we observe higher regression rate and lower mean fixation distance among successful bug location attempts in our study, 2) success or failure is not necessarily explained by having seen a correct bug line but may be better explained by regression rate, and 3) participants did not necessarily report an answer immediately after reading a correct bug line and instead read and reread other areas of code.}
\end{highlightbox}
We view these findings as evidence that visual attention differs during success or failure, and these findings serve as motivation for future work.

\section{Discussion / Conclusion}
\label{sec:conclusion}

\subsection{Contributions}
This paper makes the following key contributions:

\begin{enumerate}
    \item We perform a study of human visual attention during bug localization, specifically for memory bugs in C programs. To our knowledge, this is the first study to examine human visual attention during memory bug localization in C programs.
    \item We describe characteristics of visual attention such as the number of functions and lines to which attention is paid, highlighting specific areas of code needed to make a bug localization decision.
    \item We find preliminary evidence pointing to differences in attention associated with task success or failure.  This evidence opens the door for future work on potential positive interventions.
    \item We release all data including raw eye-tracking observations, post-processed token-level attention, etc. for future researchers.
\end{enumerate}

\subsection{Intellectual Merit}
The intellectual merit/novelty of this paper is that it is a study of C memory bugs. 
A significant majority of both program comprehension and debugging eye-tracking studies utilize Java programs~\cite{Obaidellah2018,grabinger2024eye}. The debugging eye-tracking studies that use C programs do not focus on memory bugs~\cite{li2021task,zeng2022attention,sharif2012eye,hou2013exploring}. Conducting a study specifically on C memory bugs is important because memory bugs are one of the most common programming errors in C programs~\cite{van2012memory,chromium_memory_safety}. Furthermore, C is in the top 4 most popular programming languages~\cite{he2025comparative}. To our knowledge, this  first study to examine human visual attention during memory bug localization in C programs provides information about a non-studied specific type of problem, allowing for further comparing and contrasting with other C debugging studies and well-studied problems like Java method comprehension.

\subsection{Implications}
\label{subsec:implications}

\textbf{\ref{finding:75pct} Programmers spend 75\% of their visual effort on 25\% of the functions that they view.} While the fact that some functions are more important than others when debugging is intuitive, quantifying how many functions are most important \textcolor{black}{fills a gap in the literature. As mentioned in Section~\ref{subsec:background_bug}, how visual effort is distributed in different tasks is not a settled question.} This finding also provides information that can be used in context-aware neural models and ``agentic'' AI. Specifically, these numbers provide guidance on the amount of code context that humans require to localize a bug, and therefore may inform how much context should be provided to a neural model to reduce tradeoffs. 
\textcolor{black}{Furthermore, practitioners of C programming may be interested in training interventions or tools that help them narrow their focus to a small number of methods more quickly. \textbf{Key Finding 1} may be used to inform how many visual cues tools like GazePrinter should provide~\cite{kuang2026gazeprinter,cheng2022collaborative}. These tools provide gaze-orienting visual cues to novices, nudging them to look at specific areas of code~\cite{kuang2026gazeprinter,cheng2022collaborative}.} Researchers in accessibility tools for visual impairments may also find this quantitative number informative when determining how much data should be prioritized.

\textbf{\ref{finding:fail} Even during tasks with “\textcolor{black}{location accuracy}” of 1, 2 or 3, participants often fixated on correct bug lines at least briefly.} \textcolor{black}{This means that even though participants looked at the buggy line, they did not recognize it as such. One reason for this may be that the participant had an incorrect mental model.} This finding provides fodder for a future study that investigates why participants dismiss the correct buggy line even after fixating on it. \textcolor{black}{Future studies could follow the think-aloud protocol to learn more about programmers' mental models at the time of fixation.}

\textbf{\ref{finding:statsig} As ``\textcolor{black}{Location Accuracy}'' increases, the mean euclidean distance tends to decrease, and the regression rate tends to increase.} \textbf{\ref{finding:timeline}  In
success cases, participants often spend more time
on the buggy line or regress back to the buggy
line several times.} \textcolor{black}{Together, these findings may indicate that successful participants are more ``focused''; their attention is focused in smaller regions and they re-read the same lines. This may be related to previous research finding that more successful participants had less callgraph coverage~\cite{mcloughlin2025programmers} and after a certain point, looking at more source code does not improve code summary quality~\cite{Wallace2025Programmer}.} Knowing the differences between success and failure cases may aid designers of educational tools or AI tools.  They may use these findings to inform how to detect when interventions may be needed based on eye-tracking data. This may help provide targeted instruction to students or trigger an additional training or code review in professional settings. ~\textcolor{black}{For example, if a complex section of code was reviewed without regressions, this may indicate that a more thorough review is needed. In tools like GazeCopilot, the eye-tracking data is directly integrated into  AI prompts to adapt AI suggestions to developers' cognitive states~\cite{elfares2025gazecopilot}. Future studies may also investigate which tokens are being regressed to and why.} 

\textcolor{black}{Together, the four key findings may show that successful participants strategically narrow their attention. While they may briefly skim many functions/lines, they narrow their focus to the most important 25\%. Successful participants re-read critical sections of code (regressions) and read code that is spatially near what they just finished reading (smaller euclidean distance). Future research may investigate how people perform the narrowing process, how they discover if they have narrowed down incorrectly, and how they revise their approach after such a discovery. To increase the generalizability  of these future studies, researchers should examine both dynamic and static approaches.}

\textcolor{black}{\textbf{Dataset in Aggregate:} Previous studies have used eye-tracking data to improve AI models and simulate human attention~\cite{zhang2025codeact,zhang2025eyemulator,zhang2025enhancing,zhang2024eyetrans}. This paper presents a new dataset for these purposes. Already, our data is being used to investigate the predictive power of semantic neighborhood density on programmer gaze time~\cite{wallace2026semantic}. We expect this data to be a valuable resource for additional studies.}


To encourage reproducibility of our results and as a resource for other researchers, we release all data via the following online appendix:
\url{https://github.com/apcl-research/human_attention_localize_c_bugs} (\textcolor{black}{\url{https://doi.org/10.5281/zenodo.19361519}})

\begin{acks}
This work is supported in part by the NSF grants NSF CCF-2100035 and CCF-2211428. 
Any opinions, findings, and conclusions expressed herein are the authors’ and do not necessarily reflect those of the sponsors.
\end{acks}


\end{document}